\documentclass[12pt]{article}
\usepackage{float}
\usepackage{graphicx}
\graphicspath{ {figs/} }
\usepackage{mathtools}
\usepackage{amssymb}
\usepackage{amsmath}
\usepackage{bbm}
\usepackage[inline,shortlabels]{enumitem}
\usepackage[english]{babel}
\usepackage{tikz}
\usepackage{color}
\usepackage{setspace}
\usepackage{subcaption}
\usepackage[colorlinks,citecolor=blue,urlcolor=blue]{hyperref}
\usepackage{nameref,zref-xr} 
\zxrsetup{toltxlabel} 
\usepackage{filecontents}

\makeatletter
\@ifclassloaded{biom}{%
  \usepackage{amsfonts}
  \usepackage{bm}
  \usepackage{multicol}
  \usepackage[utf8]{inputenc}
  \usepackage{standalone}
  \usepackage[figuresright]{rotating}
  \usepackage{xtab}
  \usepackage{verbatim}
  \usepackage{latexsym}
  \usepackage{booktabs}
  \usepackage{xr-hyper}
  \setcounter{footnote}{2}

  \newcommand{\titlepaper}{
    \title[Efficient nonparametric inference on the effects of stochastic
    interventions under two-phase sampling]{Efficient nonparametric inference
    on the effects of stochastic interventions under two-phase sampling, with
    applications to vaccine efficacy trials}
  }

  \newcommand{\authorlist}{
    Nima S.~Hejazi \emailx{\texttt{nhejazi@berkeley.edu}} \\
    Graduate Group in Biostatistics, University of California, Berkeley \\
    Center for Computational Biology, University of California, Berkeley
    \and
    Mark J.~{van der Laan} \emailx{\texttt{laan@berkeley.edu}} \\
    Division of Epidemiology \& Biostatistics, School of Public Health,
      University of California, Berkeley \\
    Department of Statistics, University of California, Berkeley
    \and
    Holly E.~Janes \emailx{\texttt{hjanes@fredhutch.org}} \\
    Vaccine \& Infectious Disease Division, Fred Hutchinson Cancer Research
      Center \\
    Public Health Sciences Division, Fred Hutchinson Cancer Research Center \\
    Department of Biostatistics, University of Washington
    \and
    Peter B.~Gilbert \emailx{\texttt{pgilbert@scharp.org}} \\
    Vaccine \& Infectious Disease Division, Fred Hutchinson Cancer Research
      Center \\
    Public Health Sciences Division, Fred Hutchinson Cancer Research Center \\
    Department of Biostatistics, University of Washington
    \and
    David C.~Benkeser \emailx{\texttt{benkeser@emory.edu}} \\
    Department of Biostatistics \& Computational Biology, Emory University
  }
}{%
  \usepackage{arxiv}
  \usepackage{amsthm}
  \usepackage[round]{natbib}
  \usepackage{longtable}
  \usepackage{multirow}
  \usepackage{dsfont}
  \usepackage[OT1]{fontenc}
  \usepackage{refcount}
  \usepackage[titletoc,title]{appendix}

  \newtheorem{theorem}{Theorem}
  \AtEndDocument{\refstepcounter{theorem}\label{finalthm}}
  \AtEndDocument{\refstepcounter{equation}\label{finaleq}}
  {\theoremstyle{definition}}
  {\theoremstyle{definition}}
  {\theoremstyle{definition}}

  \newtheorem{lemma}{Lemma}
  \AtEndDocument{\refstepcounter{lemma}\label{finallemma}}

  \newcommand{\titlepaper}{Efficient nonparametric inference on the effects of
    stochastic interventions under two-phase sampling, with applications to
    vaccine efficacy trials}

  \newcommand{\authorlist}{
    Nima S.~Hejazi \\
    Graduate Group in Biostatistics, and \\
    Center for Computational Biology, \\
    University of California, Berkeley \\
    \texttt{nhejazi@berkeley.edu} \\
    \And
    Mark J.~{van der Laan} \\
    Division of Epidemiology \& Biostatistics, and \\
    Department of Statistics,\\
      University of California, Berkeley \\
    \texttt{laan@berkeley.edu} \\
    \And
    Holly E.~Janes \\
    Vaccine \& Infectious Disease Division, and \\
    Public Health Sciences Division, \\
    Fred Hutchinson Cancer Research Center \\
    \texttt{hjanes@fredhutch.org} \\
    \And
    Peter B.~Gilbert \\
    Vaccine \& Infectious Disease Division, and \\
    Public Health Sciences Division, \\
    Fred Hutchinson Cancer Research Center \\
    \texttt{pgilbert@scharp.org} \\
    \And
    David C.~Benkeser \\
    Department of Biostatistics \& Bioinformatics, \\
    Rollins School of Public Health, \\
    Emory University \\
    \texttt{benkeser@emory.edu} \\
  }
}
\makeatother

\newcommand{\E}{\mathbb{E}}

\newcommand{\M}{\mathcal{M}}
\newcommand{\R}{\mathbb{R}}

\newcommand{\prob}{\mathbb{P}}

\DeclareMathOperator{\expit}{expit}
\DeclareMathOperator{\bern}{Bernoulli}
\DeclareMathOperator{\logit}{logit}

\DeclareMathOperator*{\argmin}{\arg\!\min}

\newcommand\indep{\protect\mathpalette{\protect\independenT}{\perp}}
\def\independenT#1#2{\mathrel{\rlap{$#1#2$}\mkern2mu{#1#2}}}

\zexternaldocument*[supp:]{sm}
\date{\today}
\title{\titlepaper}
\author{\authorlist}
\begin{document}
\maketitle
\begin{abstract}

The advent and subsequent widespread availability of preventive vaccines has
altered the course of public health over the past century. Despite this success,
effective vaccines to prevent many high-burden diseases, including HIV, have
been slow to develop. Vaccine development can be aided by the identification of
immune response markers that serve as effective surrogates for clinically
significant infection or disease endpoints. However, measuring immune response
is often costly, which has motivated the usage of two-phase sampling for immune
response sampling in clinical trials of preventive vaccines. In such trials,
measurement of immunological markers is performed on a subset of trial
participants, where enrollment in this second phase is potentially contingent on
the observed study outcome and other participant-level information. We propose
nonparametric methodology for efficiently estimating a counterfactual parameter
that quantifies the impact of a given immune response marker on the subsequent
probability of infection. Along the way, we fill in a theoretical gap pertaining
to the asymptotic behavior of nonparametric efficient estimators in the context
of two-phase sampling, including a multiple robustness property enjoyed by our
estimators. Techniques for constructing confidence intervals and hypothesis
tests are presented, and an open source software implementation of the
methodology, the \texttt{txshift} \texttt{R} package, is introduced. We
illustrate the proposed techniques using data from a recent preventive HIV
vaccine efficacy trial.

\end{abstract}







\section{Introduction}\label{intro}

Ascertaining the population-level causal effects of exposures is often
a motivating goal in scientific research. Such effects are commonly formulated
via summaries of the distribution of \emph{counterfactual random variables},
which describe the values a measurement would have taken if, possibly
counter-to-fact, a particular level of exposure were assigned to the unit.
Often, the exposure of interest is continuous-valued --- for example, the dose
of a drug, amount of weekly exercise, or level of an immune response marker
induced by a vaccine. The latter serves as our motivating example as we consider
data generated by a phase IIb trial of a vaccine to prevent infection by human
immunodeficiency virus (HIV), the HIV Vaccine Trials Network's (HVTN) 505
efficacy trial~\citep{hammer2013efficacy}. In addition to evaluating the overall
efficacy of the vaccine, a key secondary question of the trial was to evaluate
the role of vaccine-induced immune responses in generating protective efficacy
against HIV~\citep{janes2017higher}. Identification of immune response markers
\textit{causally} related to protection is critical both for further developing
a biological understanding of the action mechanism of a vaccine's effects and
for insights to guide the development of future vaccines.

To study such relationships, it is natural to consider a dose-response curve
that summarizes participants' risk of HIV infection as a function of the level
of a particular immune response marker. A causal formulation of such
a dose-response analysis would consider a (possibly infinite) collection of
counterfactual outcomes, each representing the HIV infection risk that would
have been observed if all individuals' immune responses had been set to
a particular level. Studying how the proportion of infected individuals varies
as a function of the level of an immune response marker could provide insights
into causal mechanisms underlying the vaccine's effects. Unfortunately, several
difficulties, both theoretical and practical, arise when considering such
a dose-response approach. From a statistical perspective, nonparametric
estimation and inference on the causal dose-response curve is challenging and
requires non-standard techniques~\citep{diaz2013targeted,
kennedy2017nonparametric,vdl2018cvtmle}. More importantly, such an approach may
require consideration of counterfactual variables that are scientifically
unrealistic. Namely, it may be impossible to imagine a world where every
participant exhibits high immune responses, simply due to phenotypic variability
of participants' immune systems. This calls into question the validity of
counterfactual dose-response analysis strategies that evaluate the effects of
immune response markers.

An alternative framework for assessing the causal effects of continuous-valued
exposures is rooted in considering counterfactual outcomes resulting from
\textit{stochastic} interventions~\citep{diaz2012population,
haneuse2013estimation, vanderweele2013causal, young2014identification}. Whereas
static interventions assign the same fixed level of an exposure to all observed
units, stochastic interventions consider instead setting the exposure level
equal to a random draw from a particular distribution. This approach provides
a more flexible means of defining counterfactual random variables. Indeed,
static interventions can be viewed as a special case of stochastic interventions
in which the intervention mechanism is drawn from a degenerate distribution with
all mass placed on only a single exposure level. In order to define
scientifically meaningful counterfactuals, care must be taken in defining the
particular distribution from which the exposure is drawn. A popular strategy is
to draw exposure levels from a modified version of the true exposure
distribution --- the \emph{natural distribution} of the exposure that occurs
under no intervention. For example, one may consider an intervention that draws
the post-vaccination level of immune response markers from a distribution that
is similar to the naturally observed (post-vaccination) distribution of immune
response markers \emph{but} that has been shifted upwards (or downwards) for
some or all participants. Counterfactuals defined by such an intervention may be
better aligned with plausible future interventions, such as refinements of the
current vaccine that provide improved immune response in some or all
participants. Evaluating the population-level risk of HIV infection under such
interventions is scientifically useful for several reasons. Firstly, this
measure of risk provides a scientifically relevant mechanism by which to
rank-order immune response markers by their importance for HIV infection risk.
Such information could be used in defining go/no-go criteria in future
early-phase HIV vaccine development pipelines. Secondly, the risk measure may
also provide a way to predict next-generation vaccine efficacy based on the
induced immune response profile, elucidating whether the immune response induced
by a candidate vaccine is sufficiently promising to advance it to the clinical
trials.


The above discussion highlights the need for rigorous methodology to identify
and estimate population-level causal effects of stochastic interventions; recent
work has provided several candidate approaches~\citep{diaz2012population,
haneuse2013estimation, young2014identification, kennedy2019nonparametric}. Both
\citet{diaz2012population} and \citet{haneuse2013estimation} propose conditions
for the identification of the causal effect under a stochastic intervention and
detail several estimators of these quantities, including strategies relying on
inverse probability weighting, outcome regression, and doubly robust estimation.
These approaches alone are insufficient for application to studies like the HVTN
505 trial, where a two-phased, case-control sampling design was used to measure
participants' immune response profiles. Under this design, all participants were
HIV-negative at the week 24 visit, with eligible cases being vaccine recipients
diagnosed with HIV-1 infection by the month 24 visit; 100\% of cases were
sampled from among those with samples available for measurement of immune
response markers at the week 26 visit, while a random sample of eligible
HIV-uninfected vaccine recipients was taken~\citep{janes2017higher}. This
sampling design severely complicates the estimation of causal effects.
\citet{rose2011targeted2sd}, among others, discuss strategies for efficient
estimation under two-phase sampling designs, emphasizing an inverse probability
of censoring weighted modification that may be coupled with targeted minimum
loss estimation (or similar frameworks) to account for study design. Their
approach yields an asymptotically linear estimator so long as the probability of
inclusion in the second-phase sample is known by design or estimable via
nonparametric maximum likelihood. This latter requirement precludes usage of
their proposed estimators in situations where, for example, sampling
probabilities depend on continuous-valued covariates. While the term ``two-phase
sampling'' has traditionally been used to denote outcome-dependent Bernoulli or
without-replacement sampling based on discrete covariates, recent efforts have
extended the concept to the usage of continuous-valued covariates in
constructing second-phase samples~\citep[e.g.,][]{chatterjee2007semiparametric,
gilbert2014optimal}. Consequently, \citet{rose2011targeted2sd} sketched a more
complicated procedure for generating efficient estimators in such settings. This
approach has neither been evaluated via simulation nor in data analysis.

In the present work, we develop estimators of the mean counterfactual outcome
under a stochastic intervention when the exposure is measured via two-phase
sampling. We provide several contributions to the disparate literatures on
two-phase sampling and stochastic interventions. To the former, we (i) formalize
the assumptions needed for efficient nonparametric inference under two-phase
sampling; (ii) characterize a multiple robustness of the estimators that arises
from the second-order remainder of a linear expansion of the target parameter;
and (iii) provide the first comparison of the practical performance of these
estimators in rigorous numerical experiments. Our contributions to the
literature on stochastic interventions are (i) a novel estimator of
a conditional density that is valid under two-phase sampling, while achieving
a fast convergence rate, a crucial development for generating efficient
estimators of the mean counterfactual; and (ii) an extension of nonparametric
inference on mean counterfactuals under stochastic interventions using
projections onto nonparametric working marginal structural models. Finally, we
provide open source software packages,
\texttt{txshift}~\citep{hejazi2020txshift} and
\texttt{haldensify}~\citep{hejazi2020haldensify}, for the \texttt{R} statistical
programming environment~\citep{R}, that facilitate implementation of our
proposed estimators and their nuisance regression functions, respectively.

The remainder of this work is organized as follows. In Section~\ref{background},
we introduce the target parameter (\ref{parameter}) and discuss prior work on
stochastic interventions (\ref{stoch_lit}) and two-phase sampling
(\ref{samp_lit}). Section~\ref{methods} includes discussion of strategies for
estimating nuisance regression functions (\ref{est_nuisance_param}) and the
formulation of two estimators (\ref{os_tml_est}) that achieve the nonparametric
efficiency bound for estimation of the counterfactual mean under a stochastic
intervention. Numerical experiments comparing the proposed estimators to simpler
variants are presented in section~\ref{sim}. Section~\ref{application} describes
an analysis of data from the HVTN 505 trial using our proposed methodology. We
conclude by summarizing the results of our statistical and scientific
investigations and discussing avenues for future investigation.

\section{Preliminaries and Background}\label{background}

\subsection{Notation, data, and target parameter}\label{parameter}

Consider data generated by typical cohort sampling: the data on a single
observational unit is denoted $X = (W, A, Y)$, where $W \in \mathcal{W}$ is
a vector of baseline covariates, $A \in \mathcal{A}$ a real-valued exposure, and
$Y \in \mathcal{Y}$ an outcome of interest. Initially, we assume access to $n$
independent copies of $X$, using $P_0^X$ to denote the distribution of $X$. We
assume a nonparametric statistical model $\mathcal{M}^X$ for $P_0^X$. We denote
by $q_{0, Y}$ the conditional density of $Y$ given $\{A, W\}$ with respect to
some dominating measure, $q_{0, A}$ the conditional density of $A$ given $W$
with respect to dominating measure $\mu$, and $q_{0, W}$ the density of $W$ with
respect to dominating measure $\nu$. We use $p_0^X$ to denote the density of $X$
with respect to the product measure. This density evaluated on a typical
observation $x$ may be expressed
\begin{equation*}\label{likelihood_factorization}
  p_0^X(x) = q_{0,Y}(y \mid A = a, W = w) q_{0,A}(a \mid W = w) q_{0,W}(w) \ .
\end{equation*}

To define a counterfactual quantity of interest, we introduce a nonparametric
structural equation model (NPSEM) to describe the data-generating process of
$X$~\citep{pearl2000causality}. Specifically, we assume the data are generated
by the following system of structural equations:
\begin{equation*}\label{npsem}
  W = f_W(U_W); A = f_A(W, U_A); Y = f_Y(A, W, U_Y),
\end{equation*}
where $\{f_W, f_A, f_Y\}$ are deterministic functions, and $\{U_W, U_A, U_Y\}$
are exogenous random variables.

The NPSEM provides a parameterization of $p_0^X$ in terms of the distribution of
the random variables $(X, U)$ modeled by the system of structural equations.
Importantly, it also implies a model for the distribution of counterfactual
random variables generated by specific interventions on the data-generating
process. For example, a \textit{static intervention} replaces $f_A$ with a value
$a$ in the support of $A$. By contrast, a \textit{stochastic intervention}
modifies the value $A$ would naturally assume, $f_A(W, U_A)$, replacing it with
a draw from a post-intervention distribution $\tilde{q}_{0,A}(\cdot \mid W)$,
where the zero subscript is included to emphasize that this distribution may
depend on $P_0^X$. A static intervention may be viewed as a particular type of
stochastic intervention in which $\tilde{q}_{0,A}(\cdot \mid W)$ places all mass
on a single point. \citet{diaz2012population} described a stochastic
intervention that draws $A$ from a distribution such that $\tilde{q}_{0,A}(a
\mid W) = q_{0,A}(a - \delta(W) \mid W)$ for a user-supplied shifting function
$\delta(W)$ and a given $a \in \mathcal{A}$. \citet{haneuse2013estimation}
showed that estimating the effect of this intervention is equivalent with that
of an intervention that modifies the value $A$ would naturally assume according
to a regime $d(A,W)$. Importantly, the regime $d(A,W)$ may depend on both the
covariates $W$ and the exposure level $A$ that would be assigned in the absence
of the regime; consequently, this has been termed a \textit{modified treatment
policy} (MTP). Both \citet{haneuse2013estimation} and \citet{diaz2018stochastic}
considered an MTP of the form $d(a,w) = a + \delta(w)$ for $\delta(w) = \gamma
\in \R$ if $a + \gamma \leq u(w)$ and $d(a,w) = a$ if $a + \gamma > u(w)$, where
$u(w)$ is the maximum value in the support of $q_{0,A}(\cdot \mid W = w)$. This
intervention generates a counterfactual random variable $Y_{A + \delta(W)}
\coloneqq f_Y(A + \delta(W), W, U_Y)$ whose distribution we denote
$P_0^{\delta}$; we seek to estimate $\psi_{0, \delta} \coloneqq
\E_{P_0^{\delta}} \{ Y_{A + \delta(W)} \}$, the mean of this counterfactual
outcome.


In the context of the HVTN 505 trial, this parameter corresponds to the
counterfactual risk of HIV-1 infection had the levels of immune response markers
of vaccinated participants been increased by $\gamma$ units relative to the
level induced by the current vaccine. This quantity may reflect the immune
response marker distribution of a next-generation HIV vaccine with improved
immunogenicity relative to the vaccine evaluated in HVTN 505. While the
magnitude of shifting could generally be allowed to vary with participant
characteristics, in our analysis, we consider an intervention that uniformly
shifts all participants' immune responses by $\gamma$ units, that is, $d(a,w)
= a + \gamma$ for all $a \in \mathcal{A}$. Note that for HVTN 505, the
parameter of interest is defined only for the vaccine group, making $A$ a
post-vaccination marker measuring an HIV-specific immune response. Importantly,
it is not conceivable to define the target parameter for placebo recipients
since only HIV-negative participants are enrolled in the trial and $A$ is only
defined if measured prior to HIV infection; consequently, all relevant placebo
recipients have value zero for the marker $A$, and it is not meaningful to apply
$d(a,w)$ to shift the distribution of $A$.

Analysis of the HVTN 505 trial is complicated by its two-phase sampling design,
a technique commonly used for assessing vaccine-mediated immune response in
efficacy trials~\citep{haynes2012immune}. In general two-phase sampling, we
do not observe $X$ on all participants. Instead, we observe $O = (W, C, C A, Y)
\sim P_0$, where $P_0$ is the distribution of $O$ and $C$ is an indicator that
an observation is included in the second-phase sample. In particular, $C_i = 1$
if $A$ is measured on the $i^{\text{th}}$ observation and $C_i = 0$ otherwise.
By convention, $C A$ denotes that unobserved values of $A$ are set to zero; this
arbitrary labeling has no effect on our subsequent developments. In the context
of vaccine trials, the probability of inclusion in the second-phase sample often
depends on $W$ and $Y$; for each $w$ and $y$ we define $g_{0, C}(y, w)
\coloneqq \prob(C = 1 \mid Y = y, W = w)$. Consequently, $P_0 \in \M
= \{P_{P^X_0, g_{0,C}}: P_0^X \in \M^X, g_{0, C}\}$ --- that is, $P_0$ is the
distribution of $O$ implied by the pair $\{P^X_0, g_{0, C}\}$. For example, in
HVTN 505 all infected participants with samples available for marker measurement
at week 28 had immune responses measured, i.e., $g_{0,C}(1,w) = 1$ for all $w$;
however, only a subset of non-infected participants had immune responses
measured. We will assume access to an i.i.d.~sample $O_1, \ldots, O_n$, denoting
its empirical distribution by $P_n$. The statistical estimation problem is thus
to develop efficient nonparametric estimators of $\psi_{0,\delta}$ based on $n$
independent realizations of the observed data unit $O$.

\subsection{Identifying the counterfactual mean under a stochastic
intervention}\label{stoch_lit}

\citet{diaz2012population} established that $\psi_{0,\delta}$ is
identified by
\begin{align}\label{eqn:identification2012}
  \psi_{0,\delta} &= \int_{\mathcal{W}} \int_{\mathcal{A}}
  \overline{Q}_{0,Y}(a, w) q_{0, A}(a - \delta(W) \mid W = w)
  q_{0, W}(w) d\mu(a)d\nu(w) \nonumber \\
  &= \int_{\mathcal{W}} \int_{\mathcal{A}}
  \overline{Q}_{0,Y}(a + \delta(w), w) q_{0, A}(a \mid W = w)
  q_{0, W}(w) d\mu(a)d\nu(w)
\end{align}
where $\overline{Q}_{0,Y}(a,w) \coloneqq \E_{P_0^X}[Y \mid A = a, W = w]$, the
conditional mean of $Y$ given $A = a$ and $W = w$, as implied by $P_0^X$. For
the statistical functional given in equation~(\ref{eqn:identification2012}) to
correspond to the causal estimand of interest, several untestable assumptions
are required, including
\begin{itemize}
  \item Consistency: $Y^{a_i + \delta(w_i)}_i = Y_i$ in the event $A_i = a_i +
     \delta(w_i)$, for $i = 1, \ldots, n$;
  \item Lack of interference (stable unit treatment value): $Y^{a_i +
    \delta(w_i)}_i$ does not depend on $a_j + \delta(w_j)$ for $i \neq j$ and
    $i = 1, \ldots, n$~\citep{rubin1978bayesian,rubin1980randomization};
  \item No unmeasured confounding (strong ignorability): $A_i \indep Y^{a_i +
    \delta(w_i)}_i \mid W = w_i$, for $i = 1, \ldots, n$; and
  \item Positivity (overlap): $a_i \in \mathcal{A} \implies a_i + \delta(w_i)
    \in \mathcal{A} \mid W = w_i$ for all $w \in \mathcal{W}$ and
    $i = 1, \ldots n$.
\end{itemize}
The positivity assumption required to establish
equation~(\ref{eqn:identification2012}) is unlike that required for static or
dynamic interventions. In particular, it does not require that the
post-intervention exposure density place mass across all strata defined by $W$.
Instead, for $\overline{Q}_{0,Y}$ to be well-defined, we require that the
post-intervention exposure mechanism be bounded, i.e., $\prob_{P_0^X}
\{q_{0,A}(A - \delta(W) \mid W) / q_{0,A}(A \mid W) > 0 \} = 1$, which is
satisfied by our choice of $\delta(W)$.

\citet{diaz2012population} further provided the efficient influence function
(EIF) of $\psi_{0,\delta}$ with respect to a nonparametric model, using this
object as a key ingredient in the construction of their proposed estimators. The
EIF, evaluated on a typical full data observation $x$, is
\begin{equation}\label{eqn:eif_full}
  D^F(P_0^X)(x) = H(a, w)\{y - \overline{Q}_{0,Y}(a, w)\} +
  \overline{Q}_{0,Y}(a + \delta(w), w) - \psi_{0,\delta},
\end{equation}
where
\begin{equation}\label{eqn:aux_covar}
  H(a, w) = \mathbbm{1}(a < u(w)) \frac{q_{0, A}(a - \delta(w) \mid w)}
  {q_{0, A}(a \mid w)} + \mathbbm{1}(a + \delta(w) \geq u(w)).
\end{equation}

\subsection{Correcting for two-phase sampling}\label{samp_lit}

Since the earliest discussion of two-phase sampling
\citep{neyman1938contribution}, a rich literature on such designs has
emerged~\citep[e.g.,][]{manski1977estimation,white1982two,
flanders1991analytic,chatterjee2003pseudoscore,
breslow2009improved,breslow2009using,dai2009semiparametric,gilbert2014optimal}.
Early proposals of estimation strategies focused on parametric models of the
sampling mechanism~\citep{breslow1988logistic}, while
\citet{robins1994estimation,robins1995semiparametric} introduced the first
semiparametric estimator to incorporate inverse probability of sampling weights.
\citet{lawless1999semiparametric} proposed a discretization-based procedure that
uses stratum membership (across a small number of discrete strata) to construct
the second-phase sample. \citet{breslow1997maximum} proposed the use of maximum
likelihood estimation (MLE) when a logistic regression may be used to link the
outcome to all covariates, while \citet{breslow2003large} established the
nonparametric MLE and its asymptotic properties for this class of problem.
Subsequent proposals include pseudoscore
estimators~\citep{chatterjee2003pseudoscore}, which may accommodate the use of
continuous covariates in the sampling mechanism via kernel smoothing but require
discrete covariates in the second-phase
sample~\citep{chatterjee2007semiparametric}; re-calibration under semiparametric
regression~\citep{chen2000unified,fong2015calibration}; and targeted minimum
loss estimation~\citep{rose2011targeted2sd}.

We build on the results of~\citet{rose2011targeted2sd}, who provide a study of
nonparametric efficiency theory in two-phase sampling designs. In particular,
they provide a representation of the EIF of a target parameter of the full data
distribution $P_0^X$ when the observed data $O_1, \ldots, O_n$ are generated by
a two-phase sampling design.
Their general developments suggest that, in the present problem, the EIF
evaluated on a typical observation from the observed data $o = (w, c, c a, y)$,
is
\begin{equation}\label{eqn:eif_obs}
  D(G_0, g_{0,C}, D^F(P_0^X))(o) = \frac{c}{g_{0,C}(y, w)} D^F(P_0^X)(o) -
  \left(\frac{c}{g_{0,C}(y,w)} - 1 \right) G_0(y,w) \ ,
\end{equation}
where $D^F(P_0^X)$ is the EIF in (\ref{eqn:eif_full}), and $G_0(y,w) \coloneqq
\E_{P_0} [D^F(P_0^X)(O) \mid C = 1, Y = y, W = w]$ is the conditional mean of
$D^F(P_0^X)$ given $Y = y$ and $W = w$ in the second-phase sample.

\citet{rose2011targeted2sd} proposed two estimation strategies. The first ---
which we call the reweighted estimator --- incorporates inverse probability
weights based on known or estimated values of the second-phase sampling
probability $g_{0,C}$ to an EIF-based estimation procedure such as one-step
estimation or targeted minimum loss (TML) estimation. The estimator is shown to
be asymptotically linear and efficient in designs where the sampling mechanism
is known or can be estimated using nonparametric maximum likelihood. The second
estimator described in this proposal is for situations where the sampling design
is unknown and must be estimated using nonparametric regression, such as when
the censoring process is outside the control of investigators; however, the
authors did not provide a formal study of the theoretical properties of this
estimator nor any numerical evaluations. Owing to its complexity, examples of
this approach are extremely limited \citep[e.g.,][]{brown2014applications}. We
aim to fill in these gaps by providing formal theory establishing conditions
under which this estimator achieves asymptotic efficiency as well as numerical
studies demonstrating its performance in the context of estimating the
counterfactual mean of a stochastic intervention.

\section{Methodology}\label{methods}

We now describe the implementation of our proposed estimators. We utilize two
frameworks for estimation: the one-step
framework~\citep{pfanzagl1985contributions,bickel1993efficient} and TML
estimation~\citep{vdl2006targeted, vdl2011targeted, vdl2018targeted}. Both
estimators develop in two stages, with crucial differences appearing in the
second stage. In the first stage, we construct initial estimators of key
nuisance quantities, while in the second stage we perform a bias-correction
based on the estimated EIF. The one-step framework updates an initial
substitution estimator by adding the empirical mean of the estimated EIF. By
contrast, the TML estimation framework uses a univariate logistic tilting model
to build a targeted estimator of $\overline{Q}_{0,Y}$, which is subsequently
used to construct an updated substitution estimator.

\subsection{Estimating nuisance parameters}\label{est_nuisance_param}

Our general strategy for estimating nuisance parameters relies on first using
the entire observed data set to estimate the second-phase sampling
probabilities, $g_{0,C}$. Subsequently, inverse probability of sampling weights
based on these estimates are used to generate estimates of relevant full data
quantities using data available only on observations in the second-phase sample.
These quantities include the outcome regression $\overline{Q}_{0,Y}$, the
exposure density $q_{0,A}$, and the joint distribution of covariates and
exposure, which we denote by $Q_{0,AW}$. Finally, estimates of full data
quantities are used to estimate $G_0$, the conditional mean of the full data EIF
given $Y$ and $W$ amongst observations included in the second-phase sample.

Excepting $Q_{0,AW}$, which we estimate using an inverse probability of
sampling weighted empirical distribution, we describe both parametric and
flexible, data adaptive estimators. The data adaptive estimators are more
parsimonious with our theoretical developments, which pertain to
nonparametric-efficient estimation; nevertheless, our developments hold equally
well for parametric working models. In theorem~\ref{theo:aslintmle}, we detail
assumptions on the stochastic behavior of estimators of these nuisance functions
and relate these to the behavior of the resultant estimator of the target
parameter.

An estimator of the sampling mechanism $g_{0,C}$ could be derived from any
classification method (e.g., logistic regression), in which $\prob_{P_0}(C = 1
\mid Y, W)$ is estimated using the full sample; however, nonparametric or
semiparametric estimation may be preferable depending on the availability of
information about the two-phase sampling design.

To generate an estimate $Q_{n,AW}$ of the full data joint distribution of
$(A,W)$, we use a stabilized inverse probability weighted empirical
distribution. For a given $(a,w)$,
\begin{equation*}
  Q_{n,AW}(a,w) \coloneqq \frac{\sum_{i=1}^n \frac{C_i}
    {g_{n,C}(Y_i, W_i)} \mathbbm{1}(A_i \le a, W_i \le w)}{\sum_{i=1}^n
     \frac{C_i}{g_{n,C}(Y_i,W_i)}} \ .
\end{equation*}

To estimate $\overline{Q}_{0,Y}$, one may again use any classification or
regression model, where $Y$ is the outcome and functions of $A$ and $W$ are
included as predictors. In fitting this model, inverse probability of sampling
weights $C_i / g_{n, C} (Y_i, W_i)$ for $i=1, \ldots, n$, are included to
account for the two-phase sampling design. Any valid regression estimator may be
leveraged for this purpose, so long as the implementation of the estimator
respects the inclusion of sample-level weights; in practice, we recommend the
use of a semiparametric or nonparametric estimator. We denote by
$\overline{Q}_{n,Y}(a,w)$ the estimate evaluated on a data unit with $A = a,
W = w$.

The simplest strategy for estimating the generalized propensity score $q_{0,A}$
is to assume a parametric working model and use standard regression techniques
to generate suitable estimates of the density. For example, one could operate
under the working assumption that $A$ given $W$ follows a Gaussian distribution
with homoscedastic variance and mean $\sum_{j=1}^p \beta_j \phi_j(W)$, where
$\phi = (\phi_j : j)$ are user-selected basis functions and $\beta = (\beta_j
: j)$ are unknown regression parameters. In this case, a density estimate would
be generated by fitting a linear regression of $A$ on $\phi(W)$ (e.g.,
minimizing inverse probability of sampling weighted least squares) to estimate
the conditional mean of $A$ given $W$, paired with an estimate of the variance
of $A$. In this case, the estimated conditional density is given by the density
of a Gaussian distribution evaluated at these estimates.

The relative dearth of available estimators of a conditional density motivated
our development of a novel estimator that accounts for two-phase sampling
designs. We detail this approach in the \href{sm}{Supplementary Materials}
and provide an implementation of our proposal in the \texttt{haldensify}
\texttt{R} package~\citep{hejazi2020haldensify}. Going forward, we denote by
$q_{n,A}(a \mid w)$ the estimated conditional density of $A$ given $W = w$,
evaluated at $a \in \mathcal{A}$.

The final nuisance parameter that must be estimated is $G_0$, the conditional
mean of the random variable $D^F(P_0^X)(O)$ given $(Y,W)$ amongst those included
in the second-phase sample. To estimate this quantity, we generate
a pseudo-outcome as follows. First, define the substitution estimator,
\begin{equation}\label{plugin}
  \psi_{n,\delta} \coloneqq \int \overline{Q}_{n,Y}(a + \delta(w), w)
  dQ_{n,AW}(a,w) = \frac{\sum_{i=1}^n \frac{C_i}{g_{n,C}(Y_i,W_i)}
  \overline{Q}_{n,Y}(A_i + \delta(W_i), W_i)}{\sum_{i=1}^n \frac{C_i}
  {g_{n,C}(Y_i,W_i)}} \ ,
\end{equation}
as well as the auxiliary term
\begin{equation*}
   H_n(a,w) \coloneqq \mathbbm{1}(a < u(w)) \frac{q_{n,A}(a - \delta(w)
   \mid w)} {q_{n,A}(a \mid w)} + \mathbbm{1}(a + \delta(w) \ge u(w)) \ .
\end{equation*}
Using these quantities, for all $i$ such that $C_i = 1$, we compute
\begin{equation*}
  D^F_{n,i} \coloneqq H_n(A_i, W_i) \{Y_i - \overline{Q}_{n,Y}(A_i, W_i)\} +
  \overline{Q}_{n,Y}(A_i + \delta(W_i), W_i) - \psi_{n,\delta} \ .
\end{equation*}
A simple estimation strategy for $G_0$ is to adopt a parametric working model
and fit, for example, a linear regression of the pseudo-outcome $D^F_{n,i}$ on
basis functions of $Y$ and $W$. Importantly, since $G_0$ is defined as
a conditional expectation with respect to the observed data distribution, we
need not include inverse probability of sampling weights in this regression
estimate. While a parametric working model for $G_0$ is permissible, given the
complexity of the object, correct specification of this model is likely
challenging and we recommend more flexible approaches. We let $G_n(Y_i,W_i)$
denote the value of the chosen regression estimator evaluated on the
$i^\text{th}$ observation $i = 1, \ldots, n$.

\subsection{Efficient estimation}\label{os_tml_est}

\subsubsection{One-step estimator}

Based on the estimated nuisance functions detailed above, efficient estimators
may be constructed using either of the one-step or targeted minimum loss
estimation frameworks. The one-step estimator adds the empirical mean of the
estimated EIF to the initial plug-in estimator,
\begin{equation}\label{eqn:one_step}
  \psi_{n,\delta}^{+} \coloneqq \psi_{n,\delta} + \frac{1}{n} \sum_{i=1}^n
   \left[\frac{C_i}{g_{n,C}(Y_i,W_i)} D^F_{n,i} -
   \left\{\frac{C_i}{g_{n,C}(Y_i,W_i)} - 1\right\} G_n(Y_i,W_i) \right] \ .
\end{equation}
The resultant augmented one-step estimator $\psi_{n,\delta}^{+}$ relies on the
nuisance functions estimators $(\overline{Q}_{n,Y}, g_{n,A}, G_n, g_{n,C})$.
Theorem~\ref{theo:aslintmle} details sufficient assumptions on these estimators
for ensuring that the one-step is asymptotically efficient.




\subsubsection{Targeted minimum loss estimator}


An asymptotically linear TML estimator of $\psi_{0,\delta}$ may be constructed
by using inverse probability of sampling weights to update the initial estimator
$\overline{Q}_{n,Y}$ to an estimator $\overline{Q}_{n,Y}^{\star}$. An updated
plug-in estimator is then constructed,
\begin{equation}\label{eqn:tmle}
  \psi_{n,\delta}^{\star} \coloneqq \int
  \overline{Q}_{n,Y}^{\star}(a + \delta(w), w) dQ_{n,AW}(a,w) \ .
\end{equation}
This updated estimator $\overline{Q}_{n,Y}^{\star}$ is constructed as follows.
\begin{enumerate}[leftmargin=1cm]
 \item Define a working logistic regression model for the conditional mean of
    $C$ given $\{Y, W\}$, $\logit(g_{n, C, \xi}) = \logit(g_{n,C}) + \xi
    (G_n / g_{n,C}) : \xi \in \R$. The MLE $\xi_n$ of the parameter $\xi$
    of this model is computed and the updated estimator $g_{n,C}^{\star}
    \coloneqq g_{n, C, \xi_n}$ is defined.
 \item Next, define a working logistic regression model for the conditional
    mean of $Y$ given $\{A,W\}$: $\logit(\overline{Q}_{n, Y, \epsilon}) =
    \logit(\overline{Q}_{n,Y})+ \epsilon H_n : \epsilon \in \mathbb{R}$. An
    estimate $\epsilon_n$ of $\epsilon$ is obtained using weighted logistic
    regression with weights $C_i / g^{\star}_{n,C}(Y_i,W_i)$ and
    $\overline{Q}_{n,Y}^{\star} \coloneqq \overline{Q}_{n, Y, \epsilon_n}$.
\end{enumerate}

The outlined procedure includes two targeting steps. The first of these steps
constructs an update of the initial estimator of the second-phase sampling
probability, $g_{n,C}^{\star}$, based on the initial estimate of $G_n$. This
step ensures that the revised estimate satisfies $\sum_{i=1}^n \{G_n(Y_i, W_i)
/ g_{n,C}^{\star}(Y_i,W_i\} \{C_i - g_{n,C}^{\star}(Y_i, W_i)\} = 0$ in a single
step when a universal least favorable submodel~\citep{vdl2016one} is used,
though an iterative procedure may be used to achieve the same result. In the
second step, the updated outcome regression $\overline{Q}_{n,Y}^{\star}$ is
generated based on the conditional density estimate $q_{n,A}$; the inclusion of
weights in the regression ensures that $\sum_{i=1}^n C_i/g_{n,C}(Y_i, W_i)
D_{n,i}^F = 0$.

When the first step of this procedure is omitted, the resultant TML estimator is
equivalent to the reweighted estimator of~\citet{rose2011targeted2sd}. The
additional step allows our estimator to attain asymptotic linearity in a broader
setting. That is, while the reweighted estimator requires that the sampling
weights be known or be estimable at a parametric rate, our approach allows for
the use of more flexible estimators of sampling weights.

\subsubsection{Asymptotic analysis of efficient estimators}\label{asymp_analy}

We establish the asymptotic efficiency of our estimators in
theorem~\ref{theo:aslintmle}. The theorem depends on a several regularity
conditions, which are discussed in the \href{sm}{Supplementary Materials}. The
theorem is provided in the context of the TML estimator, but, with a similar set
of assumptions, the same result holds for the one-step estimator; for brevity,
we omit this analogous theorem. In the sequel, $D^F(\overline{Q}_{0,Y},
q_{0,A})$ and $D^F(P_0^X)$ are used interchangeably since $D^F$ depends on
$P_0^X$ primarily through estimates of the nuisance functions
$\overline{Q}_{0,Y}$ and $q_{0,A}$.

\begin{theorem}[Asymptotic linearity and efficiency of the TML estimator
  $\psi_{n,\delta}^{\star}$]\label{theo:aslintmle} Assuming
  conditions~\ref{supp:ass:eif_rootn}-\ref{supp:ass:donsker},
  \begin{equation*}
      n^{1/2}\left(\psi_{n,\delta}^{\star} - \psi_{0,\delta}\right) =
      n^{-1/2} \sum^{n}_{i = 1} D(G_0, g_{0,C},
      D^F(\overline{Q}_{0,Y},g_{0,A}))(O_i) + o_p(1) \ .
  \end{equation*}
\end{theorem}

An immediate corollary of theorem~\ref{theo:aslintmle} is that
$\psi_{n,\delta}^{\star}$ is asymptotically efficient, since it is an
asymptotically linear estimator with influence function equal to the efficient
influence function. Moreover, the central limit theorem implies that the scaled,
centered estimator converges in distribution to a mean-zero Gaussian random
variable with variance matching that of the EIF (i.e., $\E_{P_0} \{D(G_0,
g_{0,C}, D^F(\overline{Q}_{0,Y},g_{0,A}))(O)^2\}$).

The proof of theorem~\ref{theo:aslintmle} is given in the
\href{sm}{Supplementary Materials}. The conditions of the theorem are
standard in semiparametric inference problems, essentially requiring
a sub-parametric rate of convergence of each of the nuisance estimators to their
true counterparts, a Donsker class condition on the EIF evaluated at the
estimated nuisance parameters, and $L^2(P_0)$-consistency of this same object.

With respect to this first condition, we note that the highly adaptive lasso
(HAL) regression estimator has been shown to achieve a sufficiently fast rate of
convergence so as to satisfy the requirements of the theorem
\citep{vdl2017generally, benkeser2016highly, bibaut2019fast, coyle2019hal9001}
under the assumption that the true regression function is right-hand continuous
with left-hand limits and bounded sectional variation norm. This provided
further motivation for our development of a HAL-based conditional density
estimator. We note that our simulation studies and analysis of the HVTN
505 trial data utilize HAL to increase the applicability of our theorem.

With respect to the Donsker condition, we note that this assumption may be
avoided by using cross-validation (or cross-fitting) in estimating nuisance
parameters \citep{klaassen1987consistent,zheng2011cross,chernozhukov2016double}.
Such an estimator enjoys the same asymptotic properties as our
non-sample-splitting estimator while eschewing the Donsker class condition.

\subsubsection{Multiple robustness of efficient estimators}\label{mult_robust}

The EIF of our estimators enjoys a \textit{multiple robustness} property, which
allows our estimators to achieve consistency even in situations where certain
combinations of nuisance parameters are inconsistently estimated.

\begin{lemma}[Multiple robustness of the EIF]\label{lemma:mult_robust}
  Let $(G, g_{C}, \overline{Q}_{Y}, q_{A})$ denote the limits of the nuisance
  estimators $(G_n, g^{\star}_{n,C}, \overline{Q}^{\star}_{n,Y}, q_{n,A})$ in
  probability. Suppose either of the following two conditions hold
  \begin{enumerate}[label=(\roman*)]
       \item $G = G_0$ and either $\overline{Q}_{Y} = \overline{Q}_{0,Y}$ or
          $q_{A} = q_{0,A}$;
      \item $g_{C} = g_{0,C}$ and either $\overline{Q}_{Y} = \overline{Q}_{0,Y}$
          or $q_{A} = q_{0,A}$.
   \end{enumerate}
  Then $\psi_{n,\delta}^{\star} \xrightarrow[p]{} \psi_{0,\delta}$.
\end{lemma}
In the case of the one-step estimator, the initial nuisance function estimates
$g_{n,C}$ and $\overline{Q}_{n,Y}$ are used instead. The lemma implies that our
efficient estimators will be asymptotically consistent if at least one of
$(G_0, g_{0,C})$ and one of $(\overline{Q}_{0,Y}, q_{0,A})$ are consistently
estimated.

\subsection{Confidence intervals and hypothesis tests}\label{inference}


Theorem~\ref{theo:aslintmle} established the limiting distribution of our
efficient estimators. From the limit distribution, inference for either
estimator may be attained in the form of Wald-type confidence intervals and
corresponding hypothesis tests.

Consider the null and alternative hypotheses $H_0: \psi_{0,\delta} = 0$ and
$H_1: \psi_{0,\delta} \neq 0$, and denote by $\psi_{n,\delta}$ either the TML
estimator $\psi_{n,\delta}^{\star}$ or the one-step estimator
$\psi_{n,\delta}^{+}$. Asymptotic $(1 - \alpha)$ Wald-type confidence intervals
and p-values for the hypothesis test are given by
\begin{equation*}\label{eqn:est_inference}
\text{p-value} = 2 \left[1 - \Phi \left(\frac{n^{1/2} \mid
    \psi_{n,\delta} \mid}{\sigma_n} \right) \right]
  \quad \textrm{and} \quad
  \text{CI} = \psi_{n,\delta} \pm
  z_{\left(1 - \alpha/2\right)} \sigma_n / n^{1/2} \ ,
\end{equation*}
where $\sigma^2_n$ is the empirical variance of the estimated EIF, $\Phi(\cdot)$
is the CDF of the standard normal distribution, and $z_{\left(1
- \alpha/2\right)}$ is the $1-\alpha/2$ quantile of the same distribution.

These procedures are asymptotically justified under the conditions of
theorem~\ref{theo:aslintmle}. Importantly, while multiple robustness implies
that consistent estimation of $\psi_{0,d}$ is possible under inconsistent
estimation of some nuisance parameters, the validity of confidence intervals and
hypothesis tests requires consistent estimation of \emph{all} nuisance
parameters.

\subsection{Summarization via working marginal structural
models}\label{msm_summary}

Estimation of $\psi_{0,\delta}$ for a single pre-specified shift $\delta$ may be
unsatisfactory in some contexts, as it does not provide information concerning
a dose-response relationship between exposure and outcome. Thus, to develop an
understanding of a dose-response pattern in the context of stochastic
interventions, it may be informative to estimate the counterfactual mean outcome
across several values of $\delta$. In the context of HVTN 505, we consider
estimation of a grid of counterfactual means $\psi_0 = (\psi_{n,\delta_1},
\ldots, \psi_{n,\delta_K})$ and examine how the risk of HIV infection varies
with choice of $\delta$ over a fixed grid, i.e., $\delta_k \in \{\delta_1,
\ldots, \delta_K\}$. After estimating the counterfactual mean for each
$\delta_k$, a summary measure relating the stochastic
interventions to the mean counterfactual outcomes may be constructed by
projection onto a working marginal structural model (MSM). For example, we might
consider a (possibly weighted) least-squares projection on the linear working
model $m_{\beta}(\delta) = \beta_0 + \beta_1 \delta$, in which case the
parameter $\beta_1$ corresponds to the linear trend in mean counterfactual
outcomes as a function of the $\delta_k$.

More generally, we can define $\beta(\delta)=\argmin_{\beta \in \mathbb{R}^d}
\sum_{\delta \in \{\delta_1, \ldots, \delta_K\}} h(\delta)\{\psi_{0,\delta}
- m_{\beta}(\delta)\}^2$, for a user-selected weight function $h(\delta)$. We
note that adjustment of the weight function, as well as the functional form of
$m_{\beta}(\delta)$, allow for a wide variety of working models to be
considered. Alternatively, $\beta(\delta)$ can be viewed as the solution of
\begin{equation*}
  0 = U(\beta,\psi) = \sum_{\delta \in \{\delta_1, \ldots, \delta_K\}}
  h(\delta) \frac{d}{d\beta} m_{\beta}(\delta) \{\psi_{0,\delta} -
  m_{\beta}(\delta)\}.
\end{equation*}
The goal is to make statistical inference on the parameter $\beta$. We note that
this approach does not assume a linear dose-response curve, but rather uses
a working model to summarize the relationship between exposure and outcome
\citep{neugebauer2007nonparametric}. This approach is distinct from that of
\citet{haneuse2013estimation}, whose proposal involving MSMs pertains
specifically to parametric models.

To estimate $\beta$, we assume access to the TML or one-step estimates $\psi_n
= (\psi_{n, \delta_1}, \ldots, \psi_{n,\delta_K})$ for each $\delta_k$. The
estimate $\beta_n$ of $\beta$ is the solution in $\beta$ of the equation $0
= U(\beta,\psi_n)$. To derive the limit distribution of $\beta_n$, let
$D_{0,\psi}$ denote a vector whose $k^{\text{th}}$ entry is the EIF associated
with parameter $\psi_{0,\delta_k}$. The delta method implies that the influence
function of $\beta_n$ is $D_{\beta} = [-\frac{d}{d \beta} U(\beta,\psi_0)^{-1}]
\frac{d}{d\psi_0} U(\beta, \psi_0) D_{\psi_0}$, and that $n^{1/2} (\beta_n
- \beta)$ converges in distribution to a mean-zero Gaussian random variable with
variance $\Sigma = \E_{P_0}\{D_{\beta}(O)^2\}$. The empirical covariance matrix
of $D_{0,\psi}$ evaluated at nuisance parameter estimates serves to estimate
$\Sigma$.

\section{Simulation Studies}\label{sim}

The proposed estimators were evaluated using two simulation experiments. In the
first, we compare our proposed estimators to alternative estimators proposed in
the literature. To highlight the benefits offered by our approach over the
simple reweighted estimator of \citet{rose2011targeted2sd}, we focus on how
estimation of $g_{0,C}$ influences the estimator performance. In particular, we
consider both the usage of standard logistic regression and the highly adaptive
lasso in the construction of $g_{n,C}$.

In a second simulation study, we analyze the performance of our estimators using
a data-generating mechanism that is inspired directly by data from the HVTN 505
trial. Here, we focus on comparing the relative performance of our efficient,
augmented one-step and TML estimators to assess their capabilities for use in
real-world data analysis.

In both simulation studies, we consider estimation of $\psi_{0,\delta}$ for
several values of $\delta$. These studies were performed using the
\texttt{txshift} \texttt{R} package~\citep{hejazi2020txshift}, available at
\url{https://github.com/nhejazi/txshift}.

\subsection{Simulation \#1: Comparing estimators under different sampling
mechanism estimators}\label{simple_sim}

We evaluate the estimators on data simulated from the following data-generating
mechanism:
\begin{align*}
  W_1 & \sim \mbox{Normal}(3, 1);
  W_2 \sim \bern(0.6);
  W_3 \sim \bern(0.3) \\
  A \mid W & \sim \mbox{Normal}(2(W_2 + W_3), 1) \\
  Y \mid A, W & \sim \bern \left(\expit \left((W_1 + W_2 + W_3)/3 - A \right)
    \right) \\
  C \mid Y, W & \sim \bern \left(\expit \left((W_1 + W_2 + W_3)/3 - Y
    \right) \right),
\end{align*}
where $\expit(x) = \{1 + \exp(-x) \}^{-1}$. In this setting, the outcome has a
fairly high event rate, with $\prob(Y = 1 \mid A, W) \approx 0.415$. We used
this data-generating process to sample $n$ i.i.d.~observations for $n \in
\{100, 400, 900, 1600, 2500\}$ and used the resultant data to estimate the
target parameter with each of the estimators considered. This was repeated
$1000$ times. We considered estimation of $\psi_{0, \delta}$ for $\delta \in
\{-0.5, 0, 0.5\}$, where the corresponding true values of the target parameter
were approximately $\{0.501, 0.415, 0.333\}$.

We compared the reweighted estimators of \citet{rose2011targeted2sd} to our
proposed estimators. For reference, we also present the results of a naive
estimate that ignores the two-phase sampling design. In each of these three
cases, we consider one-step and TML estimators, giving six estimators in total.
Each of these six estimators was constructed by estimating the exposure
mechanism $q_{n,A}$ and outcome mechanism $\overline{Q}_{n,Y}$ via maximum
likelihood based on correctly specified parametric models, while $g_{n,C}$ was
constructed using either logistic regression or the highly adaptive lasso.
Although a parametric regression model adequately captures the true functional
form of the sampling mechanism in this data-generating process, one may not have
access to such information about the sampling mechanism in practice. In this
case, nonparametric regression is preferable. This motivated our desire to
examine the relative performance of the reweighted versus our proposed
estimators under such a setting. Based on theory, when $g_{n,C}$ is estimated
using logistic regression, we would expect both the reweighted estimator and our
proposed estimator to be asymptotically linear, which would be supported by
observing that the bias of the estimators disappears faster than
$n^{-1/2}$-rate. On the other hand, when $g_{n,C}$ is nonparametrically
estimated (in our simulation, via the highly adaptive lasso), the reweighted
estimators should not achieve asymptotic linearity, while our proposed
estimators should. The naive estimators, which make no adjustment for the
two-phase sampling design, were expected to perform poorly --- even though this
is a ``best case'' scenario for these estimators in the sense that the two
nuisance parameters are correctly estimated using parametric models.

\begin{figure}[H]
  \centering
  \includegraphics[scale=0.35]{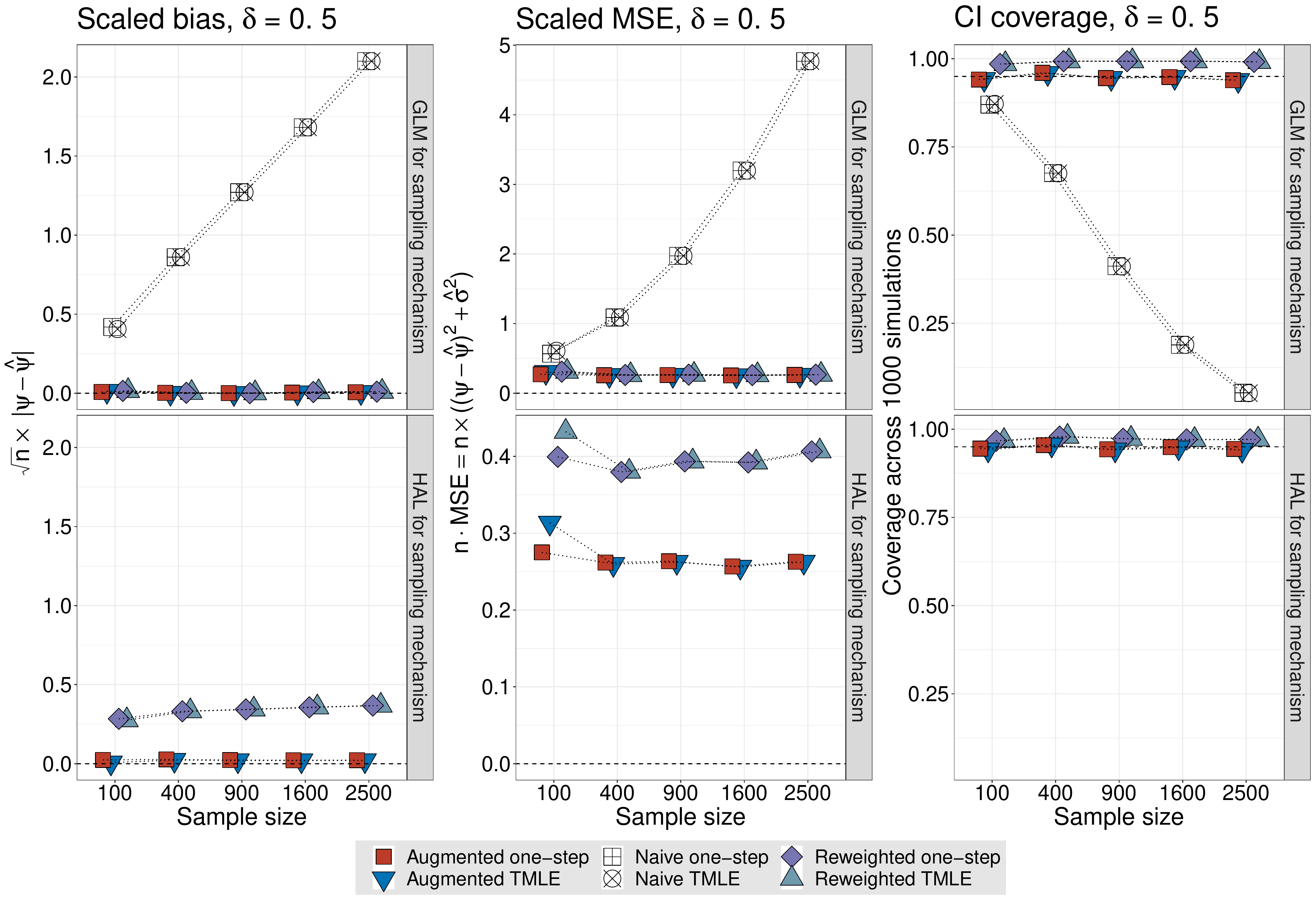}
  \caption{Results of numerical simulations comparing six estimation strategies
  for $\psi_{0,\delta}$ for $\delta = 0.5$, across 1000 Monte Carlo simulations
  for each of five sample sizes. The naive estimators do not make use of the
  estimated sampling mechanism $g_{n,C}$, so their performance is displayed
  only in the upper panel, in the interest of visual economy.}
  \label{fig:simple_sim_delta_upshift}
\end{figure}

We compared all estimators in terms of their bias (scaled by $n^{1/2})$, mean
squared-error (scaled by $n$), and coverage of 95\% Wald-style confidence
intervals. Figure~\ref{fig:simple_sim_delta_upshift} summarizes our findings for
the case $\delta = 0.5$, while Figures~\ref{supp:fig:simple_sim_delta_noshift}
and~\ref{supp:fig:simple_sim_delta_downshift}, in the
\href{sm}{Supplementary Materials}, present results for $\delta = 0$ and
$\delta = -0.5$, respectively.

When the sampling mechanism is estimated via a correctly specified parametric
model (upper panel), the reweighted and our proposed estimators behave as
expected, with low bias and stable MSE. However, the reweighted estimators
display coverage exceeding 95\%, while our proposed estimators achieve nominal
coverage. This occurs because the influence function provided in
\citet{rose2011targeted2sd}, which is the basis for the standard error estimates
used to build these confidence intervals, does not include the first-order
contribution due to estimation of the sampling mechanism. The result is
a conservative standard error estimate. Unsurprisingly, we found that the naive
estimator performed poorly in all sample sizes, highlighting the importance of
appropriately accounting for sampling design.

When the sampling mechanism was estimated using HAL (lower panel), the
reweighted estimators do not attain asymptotic linearity, as evidenced by the
scaled bias and MSE increasing with sample size. On the other hand our proposed
estimators have small bias and MSE approaching the efficiency bound, thus
demonstrating the benefits of the additional effort required to produce our
estimators over the simpler reweighted estimators.

\subsection{Simulation \#2: Comparing estimators in a scenario based on the
HVTN 505 trial}\label{hvtn_sim}

For use in real data analysis, we examine only our augmented one-step and TML
estimators. We used the HVTN data to calibrate the following data-generating
mechanism:
\begin{align*}
  W_1 & \sim \mbox{Normal}(26.6, 5.7);
  W_2 \sim \text{Poisson}(40);
  W_3 \sim \bern(0.4);
  W_4 \sim \bern(0.3) \\
  A \mid W & \sim \mbox{Normal}(-1.37 + 0.004 W_1 + 0.015 W_2 + 0.05 W_3 +
    0.25 W_4, 0.2^2)\\
  Y \mid A, W & \sim \bern \left(\expit (-2.9 - 0.0013 W_1 - 0.0016 W_2 +
    0.0678 W_3 + 0.039 W_4 - 0.033 A) \right) \\
  C \mid Y, W & \sim
    \begin{cases}
      \bern \left(\expit (-2.45 - 0.027 W_1 + 0.012 W_2 + 0.39 W_3
      + 0.166 W_4) \right), & Y = 0 \\ 1, & Y = 1
    \end{cases} \ .
\end{align*}
For consistency with the observed rate of HIV infection in the HVTN 505 trial,
the outcome was made to have a relatively rare event rate, with $\prob(Y = 1
\mid A, W) \approx 0.059$. We considered a setting where we observed $n = 1400$
i.i.d.~observations, approximately matching the sample size of the vaccinated
arm in HVTN 505. Estimator performance was assessed and reported by
aggregating across 1000 repetitions. Under this data-generating mechanism, the
true parameter values were approximated as $\psi_{0,\delta} = \{0.0627,
0.0617, 0.0609, 0.0598, 0.0589, \allowbreak 0.0580, 0.0571, 0.0561, 0.0554\}$
for $\delta \in \{-2.0, -1.5, -1.0, \allowbreak -0.5, 0.0, 0.5, 1.0, 1.5,
2.0\}$, respectively. An analogous setting, in which the effect of exposure on
the outcome was removed, yielded similar results; these are presented in the
\href{sm}{Supplementary Materials}.

\begin{figure}[H]
  \centering
  \includegraphics[scale=0.35]{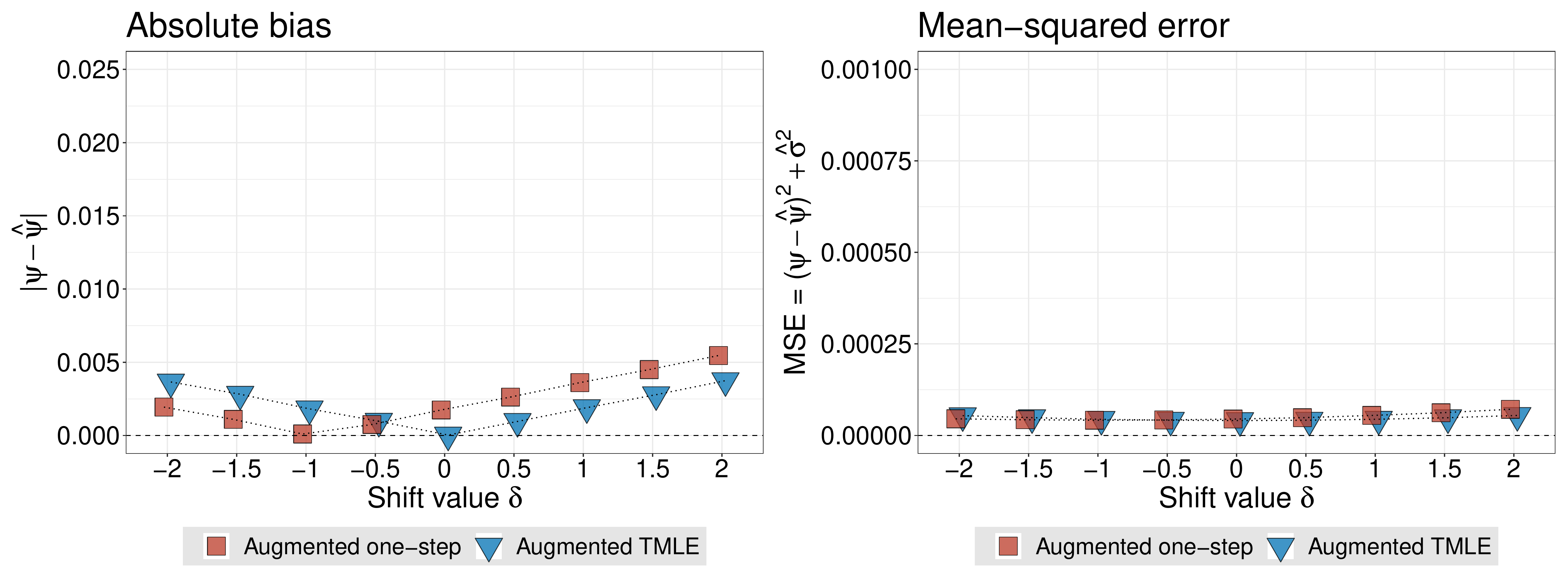}
  \caption{Results of numerical simulations comparing our two proposed
  estimators of $\psi_{0,\delta}$ for a grid of $\delta$, across 1000 Monte
  Carlo simulations at $n = 1400$.}
  \label{fig:hvtn_sim_moderate}
\end{figure}

The proposed estimators were constructed by using the highly adaptive lasso to
estimate the sampling mechanism $g_{n,C}$, the exposure mechanism $q_{n,A}$, the
outcome mechanism $\overline{Q}_{n,Y}$, and the pseudo-outcome regression $G_n$.
We compared the proposed one-step and TML estimators in terms of their bias and
MSE. The results are summarized in Figure~\ref{fig:hvtn_sim_moderate}.


Inspection of the bias and MSE indicates that both of our proposed estimators
display adequate performance, with small finite-sample bias and mean-squared
error. In terms of bias and MSE, the performance of the two estimators is nearly
indistinguishable. From these numerical investigations, we concluded that our
augmented estimators are both well-suited to estimating $\psi_{0,\delta}$ in the
context of a re-analysis of data from the HVTN 505 trial.

\section{Application to the HVTN 505 Trial}\label{application}

The HVTN 505 trial was a randomized control trial that enrolled $2504$
HIV-negative participants and randomized participants 1-to-1 to receive an
active vaccine or placebo. The one-year incidence of HIV-1 infection was about
1.8\% per person-year in the vaccine arm and 1.4\% per person-year in the
placebo arm, during primary follow-up for HIV-1 acquisition (between week 28
and month 24; the same period as was used for assessment of immune
correlates). Blood was drawn post-vaccination and immune responses measured
via intracellular cytokine staining of preserved HIV-1-stimulated peripheral
blood mononuclear cells for all HIV-1 cases diagnosed between week 28 and
month 24 and a random sample of uninfected controls~\citep{janes2017higher}.
The two-phase sampling of vaccine recipient controls without-replacement
sampled five controls per case within each of eight baseline covariate strata
defined by categories of body mass index and race/ethnicity (White, Hispanic,
Black). \citet{janes2017higher} provide the first immune correlates analysis
of the HVTN 505 data based on the constructed two-phase sample, focusing on
two cellular immune responses: the Envelope CD4+ and CD8+ polyfunctionality
scores. Subsequently, \citet{fong2018modification} analyzed an expanded set of
immune markers, including six primary humoral immune response variables beyond
the two first studied by~\citet{janes2017higher}. Both sets of analyses found
the CD4+ and CD8+ polyfunctionality scores to be informative of HIV-1
infection status by month 24 of the study (end of follow-up).

We examined how a range of posited shifts in standardized polyfunctionality
scores of the CD4+ and CD8+ immune markers would impact the mean counterfactual
risk of HIV-1 infection in vaccine recipients. We considered
\textit{standardized} polyfunctionality scores, so that our pre-specified grid
of shifts $\delta \in \{-2.0, -1.5, -1.0, -0.5, \allowbreak 0.0, 0.5, 1.0, 1.5,
2.0\}$ can be interpreted as shifts on the scale of standard deviation (sd). We
present results based on our TML estimator; results for the one-step estimator
were similar. In order to summarize the manner in which the mean counterfactual
risk of HIV-1 infection varies with shifts in the polyfunctionality scores, we
consider projection onto a linear working MSM, as discussed in
section~\ref{msm_summary}. Our augmented TML estimator $\psi_{n,\delta}^{\star}$
for the mean counterfactual risk of HIV-1 infection requires the construction of
initial estimators of all nuisance functions.

We used flexible estimation strategies for each of the nuisance parameters. The
conditional probability of inclusion in the second-phase sample was estimated
using HAL, adjusting for age, sex, race/ethnicity, body mass index, and
a behavioral risk score for HIV-1 infection. The density $q_{n,A}$ of the CD4+
or CD8+ polyfunctionality scores, conditional on the same set of covariates, was
estimated using our proposed HAL-based density
estimator~\citep{hejazi2020haldensify}. The outcome regression
$\overline{Q}_{0,Y}$ was estimated using super learner~\citep{vdl2007super},
which was used to build an ensemble model as a weighted combination of
prediction functions from a diverse collection of candidate algorithms. Further
details, including choices of learning algorithms and weights assigned to each
by the super learner ensemble, are given in section~\ref{supp:hvtn_sl_details}
of the \href{sm}{Supplementary Materials}. The pseudo-outcome regression, $G_n$,
was fit via HAL.

We note that our choice of estimation strategy --- specifically, the use of HAL
regression for $g_{0,C}$ and HAL-based conditional density estimation for
$q_{0,A}$ --- ensures a sufficiently fast rate of convergence of these nuisance
parameter estimates to their true values under only mild assumptions.
Importantly, this allows for the second case described in
lemma~\ref{lemma:mult_robust} to be fulfilled, ensuring consistency of the
resultant TML estimates. This consistency of the estimates would hold even in
the case that the HAL regression and super learner ensembles used for $G_0$ and
$Q_{0,Y}$, respectively, proved inconsistent; however, as both of these nuisance
parameters are also estimated via flexible modeling approaches, it would be
expected that the conditions of theorem~\ref{theo:aslintmle} are made to hold.
Thus, we expect that our TML estimates of the counterfactual risk of HIV-1
infection across the grid in $\delta$ will be consistent and (quite possibly)
nonparametric-efficient. Taking advantage of the robustness built into our
estimation strategy lends an additional degree of reliability as to the results
of our analysis of the HVTN 505 data. Finally, we recall that the estimate of
$\beta_0$, the slope of the marginal structural model through the TML estimates
of the counterfactual risk in $\delta$, is endowed with the same consistency and
efficiency properties as the individual TML estimates.

\begin{figure}[H]
  \begin{subfigure}{0.9\textwidth}
  \centering
  \includegraphics[scale=0.3]{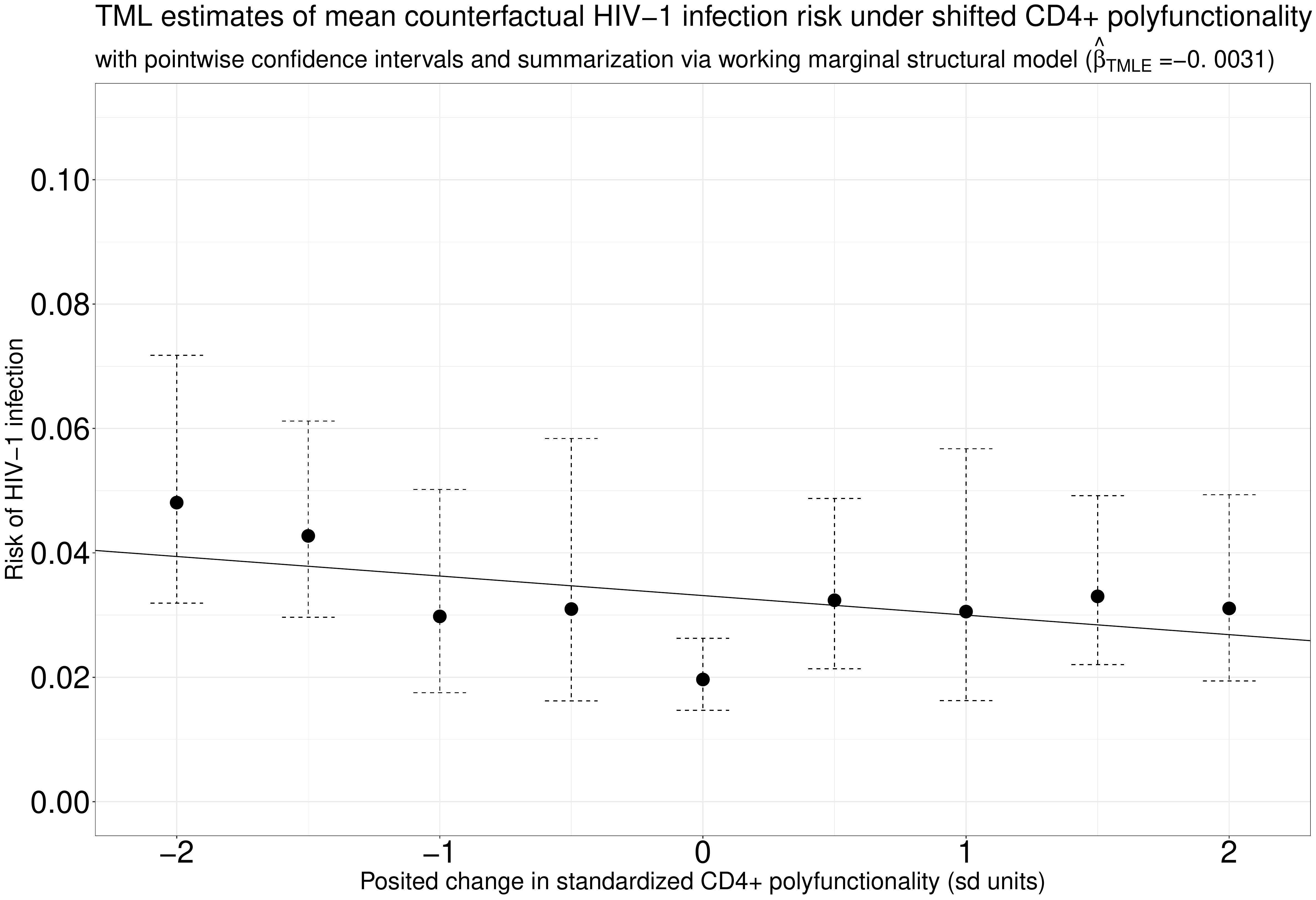}
  \label{fig:cd4_tmle_msm}
  \end{subfigure}\\[0.5cm]
  \begin{subfigure}{0.9\textwidth}
  \centering
  \includegraphics[origin=c,scale=0.3]{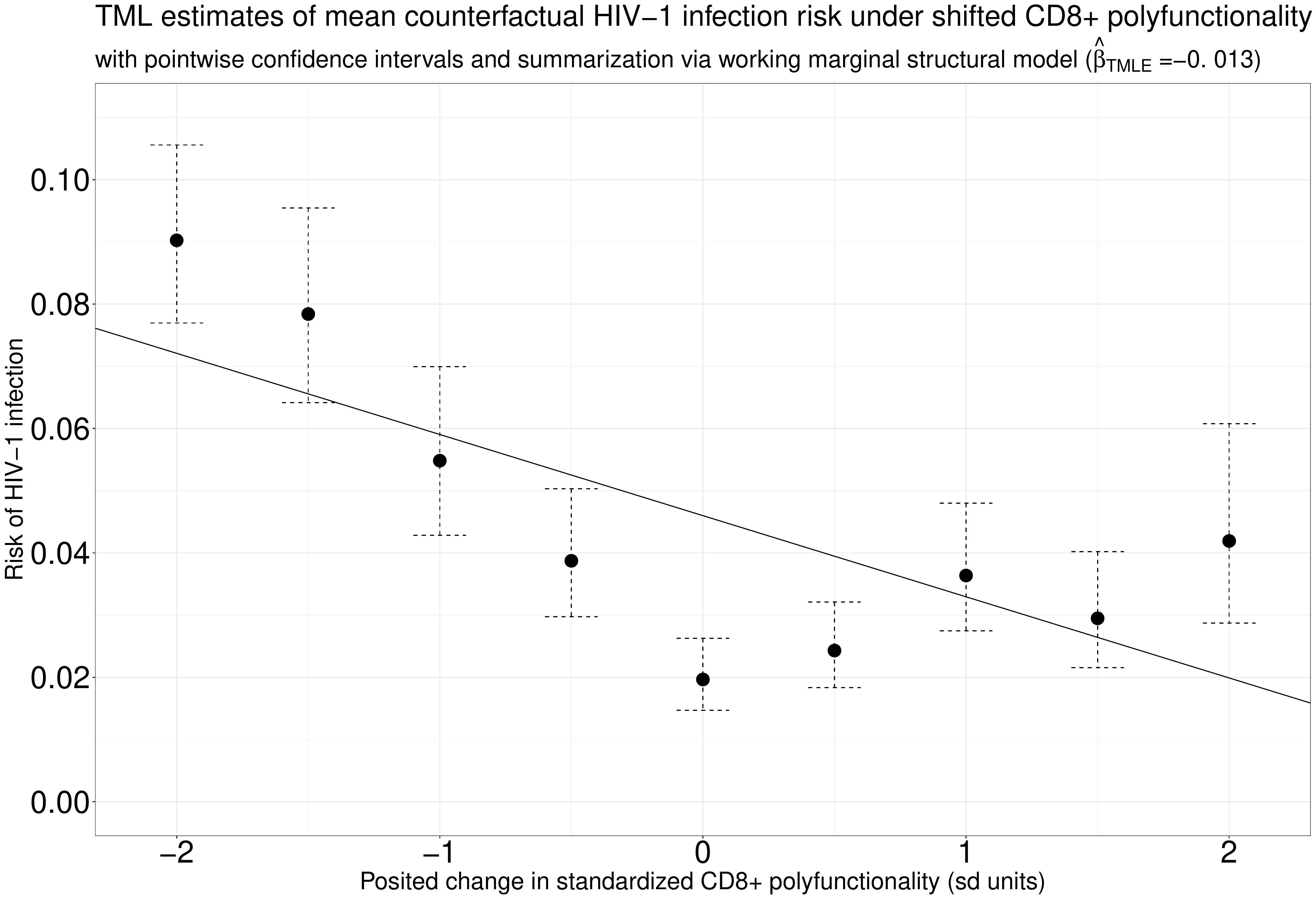}
  \label{fig:cd8_tmle_msm}
  \end{subfigure}
  \caption{TML estimates of the counterfactual mean of HIV-1 infection under
    stochastic interventions on CD4+ (top) and CD8+ (bottom) standardized
    polyfunctionality scores. Inference for the estimates is based on pointwise
    Wald-type confidence intervals. The slope of a linear working MSM
    $\hat{\beta}_{\text{TMLE}}$ summarizes the effect of shifting the
    polyfunctionality scores on the mean counterfactual risk of HIV-1
    infection.}
  \label{fig:hvtn505_tmle_msm}
\end{figure}

Results of applying our estimation procedure separately to both the CD4+ and
CD8+ polyfunctionality scores are presented in
Figure~\ref{fig:hvtn505_tmle_msm}.

Examination of the point estimates and confidence intervals of $\psi_{0,\delta}$
in Figure~\ref{fig:hvtn505_tmle_msm} reveals that downshifts in the CD4+
polyfunctionality score led to a small increase in estimated HIV-1 infection
risk (Figure~\ref{fig:hvtn505_tmle_msm}, top panel). For example, a shift of two
standard units lower in the CD4+ polyfunctionality score was found to double the
risk of HIV-1 infection. The estimated slope parameter of the working MSM
$\hat{\beta}_{\text{TMLE}}$ pointed to an estimated decrease in risk of about
-0.3\% per standard unit of CD4+ polyfunctionality change.

The estimated result of shifts in the polyfunctionality score of the CD8+
immunogenic marker displayed a markedly much stronger relationship with the risk
of HIV-1 infection (Figure~\ref{fig:hvtn505_tmle_msm}, lower panel). While
positive shifts of the standardized CD8+ polyfunctionality score beyond those
observed in the trial do not appear to have a strong effect on HIV-1 infection
risk, shifts that lower the CD8+ polyfunctionality score display a negative
linear trend, indicating that decreases in CD8+ marker activity adversely affect
the risk of HIV-1 infection. At the largest negative shift considered, the
counterfactual HIV-1 infection risk is over four times that observed in the
HVTN 505 trial.

Overall, the results of our analyses support the conclusions of
\citet{janes2017higher} and \citet{fong2018modification}, further indicating
that modulation of the CD4+ and CD8+ polyfunctionality scores may reduce the
risk of HIV-1 infection, with CD8+ polyfunctionality playing a particularly
important role. Notably, our analysis differs from the previous efforts in two
ways: our estimates (i) are based on a formal causal model, which provides an
alternative estimand to summarize relationships between immunogenic response and
risk of HIV-1 infection, and (ii) leverage  machine learning to allow the use of
flexible modeling strategies while simultaneously delivering robust inference.


\section{Discussion}\label{discuss}

Our analysis of the HVTN 505 trial could be improved in several respects. First,
there was participant dropout observed in the trial, which our analysis ignored.
A more robust analysis could leverage available covariate information to account
for potentially informative missingness.
Beyond this issue, there are several other directions for potentially
interesting extensions. It would be of interest to extend our estimation
strategy to other effects based on stochastic interventions, including the
population intervention (in)direct effects~\citep{diaz2020causal} defined in
mediation analysis and the incremental propensity score interventions
of~\citet{kennedy2019nonparametric}. Extending our estimation strategy to such
settings and its application in analyzing other vaccine efficacy trials will be
the subject of future research.




\bibliographystyle{biom}
\bibliography{refs}
\end{document}


\maketitle

\section{Conditional Density Estimation Based on the Highly Adaptive
Lasso}\label{cond_dens}

In section~\ref{main:est_nuisance_param}, several of the challenges associated
with conditional density estimation were briefly expressed. The auxiliary
covariate given in equation (\ref{main:eqn:aux_covar}) of the EIF is a ratio of
such conditional densities. In the construction of our proposed estimators, the
exposure mechanism $q_{0,A}$ is a density of the intervention $A$, conditional
on the observed covariates $W$, which must be evaluated at both the observed
values and the post-intervention counterfactual values. As consistent estimation
of the the generalized propensity score is an integral part of our proposed
methodology, we outline here a conditional density estimator, built around the
HAL regression function, that achieves the convergence rate required by our
formal theorem. We note that proposals for the data adaptive estimation of such
quantities are sparse in the literature~\citep[e.g.,][]{zhu2015boosting}.
Notably, \citet{diaz2011super} gave a proposal for constructing semiparametric
estimators of such a target quantity based on exploiting the relationship
between the hazard and density functions. Our proposal builds upon theirs in
several key ways: (i) we adjust their algorithm so as to incorporate
sample-level weights, necessary for making use of inverse probability of
censoring weights; and (ii) we replace their use of an arbitrary classification
model with one based on the HAL regression function. While our first
modification is general and may be applied to the estimation strategy
\citet{diaz2011super} propose, our latter contribution requires adjusting the
penalization aspect of HAL regression to respect the use of a loss function
appropriate for prediction on the hazard scale. To make these contributions
widely accessible, we introduce the \texttt{haldensify} \texttt{R}
package~\citep{hejazi2020haldensify}, available at
\url{https://github.com/nhejazi/haldensify}

To build an estimator of a conditional density, \citet{diaz2011super} considered
discretizing the observed $a \in A$ based on a number of bins $T$ and a binning
procedure (e.g., including the same number of points in each bin or forcing
bins to be of the same length). We note that the choice of the tuning parameter
$T$ corresponds roughly to the choice of bandwidth in classical kernel density
estimation; this will be made clear upon further examination of the proposed
algorithm. The data $\{A, W\}$ are reformatted such that the hazard of an
observed value $a \in A$ falling in a given bin may be evaluated via standard
classification techniques. In fact, this proposal may be viewed as
a re-formulation of the classification problem into a corresponding set of
hazard regressions:
\begin{align*}
   \prob (a \in [\alpha_{t-1}, \alpha_t) \mid W) =& \prob (a \in [\alpha_{t-1},
   \alpha_t) \mid A \geq \alpha_{t-1}, W) \times  \\ & \prod_{j = 1}^{t -1} \{1
   - \prob (a \in [\alpha_{j-1}, \alpha_j) \mid A
   \geq \alpha_{j-1}, W) \},
\end{align*}
where the probability that a value of $a \in A$ falls in a bin $[\alpha_{t-1},
\alpha_t)$ may be directly estimated from a standard classification model. The
likelihood of this model may be re-expressed in terms of the likelihood of
a binary variable in a data set expressed through a repeated measures structure.
Specifically, this re-formatting procedure is carried out by creating a data set
in which any given observation $A_i$ appears (repeatedly) for as many intervals
$[\alpha_{t-1}, \alpha_t)$ that there are prior to the interval to which the
observed $a$ belongs. A new binary outcome variable, indicating $A_i \in
[\alpha_{t-1}, \alpha_t)$, is recorded as part of this new data structure. With
the re-formatted data, a pooled hazard regression, spanning the support of $A$
is then executed. Finally, the conditional density estimator
\begin{equation*}
   q_{n, \alpha}(a \mid W) = \frac{\prob(a \in [\alpha_{t-1}, \alpha_t)
      \mid W)}{(\alpha_t - \alpha_{t-1})},
\end{equation*}
for $\alpha_{t-1} \leq a < \alpha_t$, may be constructed. As part of this
procedure, the hazard estimates are mapped to density estimates through
rescaling of the estimates by the bin size ($\alpha_t - \alpha_{t-1}$).

In its original proposal, a key element of this procedure was the use of any
arbitrary classification procedure for estimating $\prob(a \in [\alpha_{t-1},
\alpha_t) \mid W)$, facilitating the incorporation of flexible, data adaptive
estimators. We alter this proposal in two ways, (i) replacing the arbitrary
estimator of $\prob(a \in [\alpha_{t-1}, \alpha_t) \mid W)$ with HAL regression
and (ii) accommodating the use of sample-level weights, making it possible for
the resultant conditional density estimator to achieve a convergence rate with
respect to a loss-based dissimilarity of $n^{-1/4}$ under only mild assumptions.
This is an important advance that is needed for the asymptotic analysis of our
proposed estimators, as per section~\ref{main:asymp_analy}. As a secondary
advance, our procedure alters the HAL regression function to use a loss function
tailored for estimation of the hazard, invoking $\ell_1$-penalization in
a manner consistent with this loss.

\section{Proof of Theorem~\ref{main:theo:aslintmle}}

We now examine a proof of the theorem establishing conditions for the weak
convergence of our efficient estimators. Building upon
section~\ref{main:background}, note that the full data parameter may be
expressed as a mapping $\Psi^F: \M^X \rightarrow \R$ and that $\Psi^F(P_0^X)
\equiv \Psi^F(\overline{Q}_{0,Y})$, since the parameter mapping depends on
$P_0^X$ only through the functional $\overline{Q}_{0,Y}$. We recall that the EIF
$D^F(\overline{Q}_{0,Y},g_{0,A})$ coincides with the canonical gradient of the
parameter mapping $\Psi^F: \M^X \rightarrow \R$, since a regular asymptotically
linear estimator with influence function equal to the canonical gradient is
asymptotically efficient~\citep{bickel1993efficient, vdl2003unified}. Note
further that the observed data parameter is defined such that $\Psi(P_0) \equiv
\Psi^F(P^X_0)$, where the observed data parameter mapping $\Psi:\M \rightarrow
\R$ is pathwise differentiable at a distribution $P$ in the statistical model
with a gradient given by the EIF:
\begin{equation*}
  D(G_0, g_{0,C}, D^{F}(\overline{Q}_{0,Y}, q_{0,A}))(o) =
  \frac{C}{g_{0,C}(y, w)} D^F(\overline{Q}_{0,Y}, q_{0,A})(x) -
  \frac{G_0(y,w)}{g_{0,C}(y,w)}(C - g_{0,C}(y,w)) \ .
\end{equation*}
The class of all gradients of $\Psi$ at $P$ is given by $\{D(G_0, g_{0,C},
D^F(P^X)): D^F(P^X)\}$ where $D^F(P^X)$ varies over all gradients of the
full data parameter $\Psi^F$ at distributions $P^X \in \M^X$. As $D^F$ varies,
$G_0$ also necessarily varies since it is defined as the conditional mean of
$D^F$ given $\{C = 1, Y, W\}$. In particular, if the full data model $\M^X$ is
nonparametric, then there is only one full data gradient, which is the canonical
gradient (or EIF) of $\Psi$ at $P$.

In the sequel, let $\lVert f \rVert_2 = \E\{f(O)^2\}^{1/2}$ denote the
$L^2(P_0)$ norm of a $P_0$-measurable function $f$, define $\widetilde{G}_0
\coloneqq \E_{P_0}[D^F(\overline{Q}^{\star}_{n,Y}, q_{n,A})(O) \mid C = 1, Y,
W]$, and let $G_n$ be the estimate of $\widetilde{G}_0$. An exact
characterization of the second-order remainder for the full data parameter
$\Psi^F(P^X) - \Psi^F(P_0^X)$ is given by
\begin{equation*}
  R^F_2(\overline{Q}_{n,Y}, q_{n,A}, \overline{Q}_{0,Y}, q_{0,A}) \coloneqq
    \Psi^F(\overline{Q}_{n,Y})-\Psi^F(\overline{Q}_{0,Y}) +
    \E_{P^X_0}[D^F(\overline{Q}_{n,Y}, q_{n,A})]
\end{equation*}
while the exact second-order remainder for the observed data parameter
$\Psi(P)-\Psi(P_0)$ is analogously defined
\begin{equation*}
  R_2(P, P_0) \coloneqq \Psi(P) - \Psi(P_0) + \E_{P_0}[D(G_n, g_{n,C},
  D^F(\overline{Q}_{n,Y}, q_{n,A}))].
\end{equation*}
Combining these definitions, the exact second-order remainder for the observed
data parameter may be expressed in terms of the second-order remainder of its
full data counterpart:
\begin{equation*}
  R_2(P, P_0) = R_2^F(\overline{Q}_{n,Y}, q_{n,A}, \overline{Q}_{0,Y}, q_{0,A})
    + \E_{P_0} \left[ \left(\frac{g_{n,C} - g_{0,C}}{g_{n,C}} \right) (G_n -
    \widetilde{G}_0) \right].
\end{equation*}

Assume the following conditions:
\begin{assumption}\label{ass:eif_rootn}
  $\E_{P_n} D(G_n, g^{\star}_{n,C}, D^F(\overline{Q}^{\star}_{n,Y},
    q_{n,A})) = 0$.
\end{assumption}
\begin{assumption}\label{ass:sampling_bound}
  $g_{0,C} > \zeta > 0$ and $g^{\star}_{n,C} > \zeta$ with probability tending
  to 1 for some $\zeta > 0$.
\end{assumption}
\begin{assumption}\label{ass:obs_eif_parametric}
    $\lVert G_n - \widetilde{G}_0 \rVert_{P_0} = o_P(n^{-1/4})$ and
        $\lVert g^{\star}_{n,C} - g_{0,C} \rVert_{P_0} = o_P(n^{-1/4})$.
\end{assumption}
\begin{assumption}\label{ass:r2_fulldata}
  $R_2^F(\overline{Q}^{\star}_{n,Y}, q_{n,A}, \overline{Q}_{0,Y},
       q_{0,A}) = o_P(n^{-1/2})$.
\end{assumption}
\begin{assumption}\label{ass:neglig_obs_eif}
   $\lVert D(G_n, g^{\star}_{n,C}, D^F(\overline{Q}^{\star}_{n,Y},
        q_{n,A})) - D(G_0, g_{0,C}, D^{F}(\overline{Q}_{0,Y}, q_{0,A}))
        \rVert_{P_0} = o_P(1)$.
\end{assumption}
\begin{assumption}\label{ass:donsker}
   Let $\mathcal{F}_v^{\star}(M)$ be the class of cadlag functions $f$
   on a cube $[0,\tau] \subset \R^d$ (for some integer $d$), for which the
   sectional variation norm $\lVert f \rVert_v^{\star}$ is bounded by a
   universal constant $M < \infty$. Assume that $D(G_n,
   g^{\star}_{n,C}, D^{F}(\overline{Q}^{\star}_{n,Y}, q_{n,A})) \in
   \mathcal{F}_v^{\star}(M)$ with probability tending to 1 (n.b., the
   definition $\mathcal{F}_v^{\star}(M)$ can be replaced by any Donsker class).
\end{assumption}

\begin{proof}[Theorem~\ref{main:theo:aslintmle}: asymptotic linearity and
efficiency of the TML estimator $\psi_{n,\delta}^{\star}$]\label{pf:aslintmle}
Under conditions~\ref{ass:eif_rootn}--\ref{ass:donsker}, we have $\E_{P_n}
D(G_n, g^{\star}_{n,C}, D^F(\overline{Q}^{\star}_{n,Y},q_{n,A})) =
o_P(n^{-1/2})$; moreover, by definition of $R_2(P, P_0)$:
\begin{align*}
  \Psi^F(\overline{Q}^{\star}_{n,Y}) - \Psi^F(\overline{Q}_{0,Y}) =
   &-\E_{P_0} [D(G_n, g^{\star}_{n,C},
      D^F(\overline{Q}^{\star}_{n,Y}, q_{n,A}))] \\
   &+ R^F_2(\overline{Q}^{\star}_{n,Y}, q_{n,A}, \overline{Q}_{0,Y}, q_{0,A})
  + \E_{P_0} \left[ \left( \frac{g^{\star}_{n,C} - g_{0,C}}
   {g^{\star}_{n,C}} \right) (G_n - \widetilde{G}_0)\right].
\end{align*}
Combining with condition~\ref{ass:eif_rootn}, we have
\begin{align*}
  \Psi^F(\overline{Q}^{\star}_{n,Y}) - \Psi^F(\overline{Q}_{0,Y}) =
  & \E_{P_n}[D(G_n, g^{\star}_{n,C}, D^F(\overline{Q}^{\star}_{n,Y}, q_{n,A}))]
  - \E_{P_0}[D(G_n, g^{\star}_{n,C}, D^F(\overline{Q}^{\star}_{n,Y},
  q_{n,A}))] \\
  &+ R^F_2(\overline{Q}^{\star}_{n,Y}, q_{n,A}, \overline{Q}_{0,Y}, q_{0,A}) +
  \E_{P_0} \left[ \left( \frac{g^{\star}_{n,C} - g_{0,C}}{g^{\star}_{n,C}}
  \right) (G_n - \widetilde{G}_0) \right]
\end{align*}

Conditions~\ref{ass:neglig_obs_eif} and~\ref{ass:donsker} imply that the sum of
the first two terms equals $\E_{P_n}[D(G_0, g^{\star}_{0,C},
D^F(\overline{Q}^{\star}_{0,Y}, q_{0,A})(O)] + o_{P}(n^{-1/2})$.
Condition~\ref{ass:r2_fulldata} ensures the third term equals $o_{P}(n^{-1/2})$.
Conditions~\ref{ass:sampling_bound} and~\ref{ass:obs_eif_parametric} imply that
the fourth term equals $o_P(n^{-1/2})$, which completes the proof. An analogous
proof holds for the one-step estimator $\psi_{n,\delta}^{+}$ when the initial
estimates $\overline{Q}_{n,Y}$ and $g_{n,C}$ are used instead of their revised
counterparts. Here, we do not require condition~\ref{ass:eif_rootn}, as our
proposed estimation procedure guarantees that it will be attained.
\end{proof}

\section{Results of Additional Simulation Studies}\label{more_sims}

\subsection{Simulation \#1: Comparison of estimator variants under the
shifts $\delta = -0.5$ and $\delta = 0$}\label{sim1_supp}

\begin{figure}[H]
  \centering
  \includegraphics[scale=0.33]{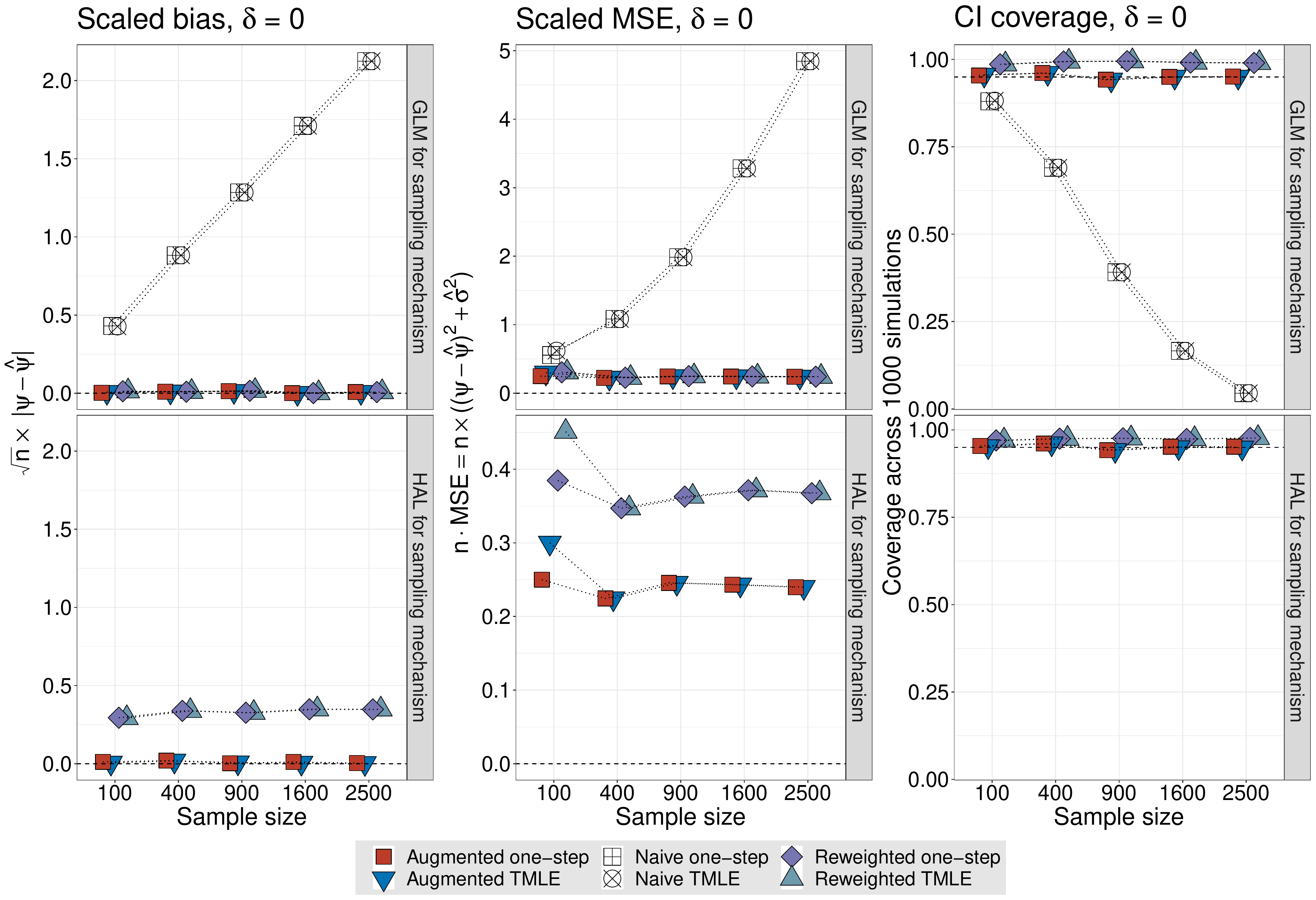}
  \caption{Results of numerical simulations comparing six estimation strategies
  for $\psi_{0,\delta}$ for $\delta = 0$.}
  \label{fig:simple_sim_delta_noshift}
\end{figure}

\begin{figure}[H]
  \centering
  \includegraphics[scale=0.33]{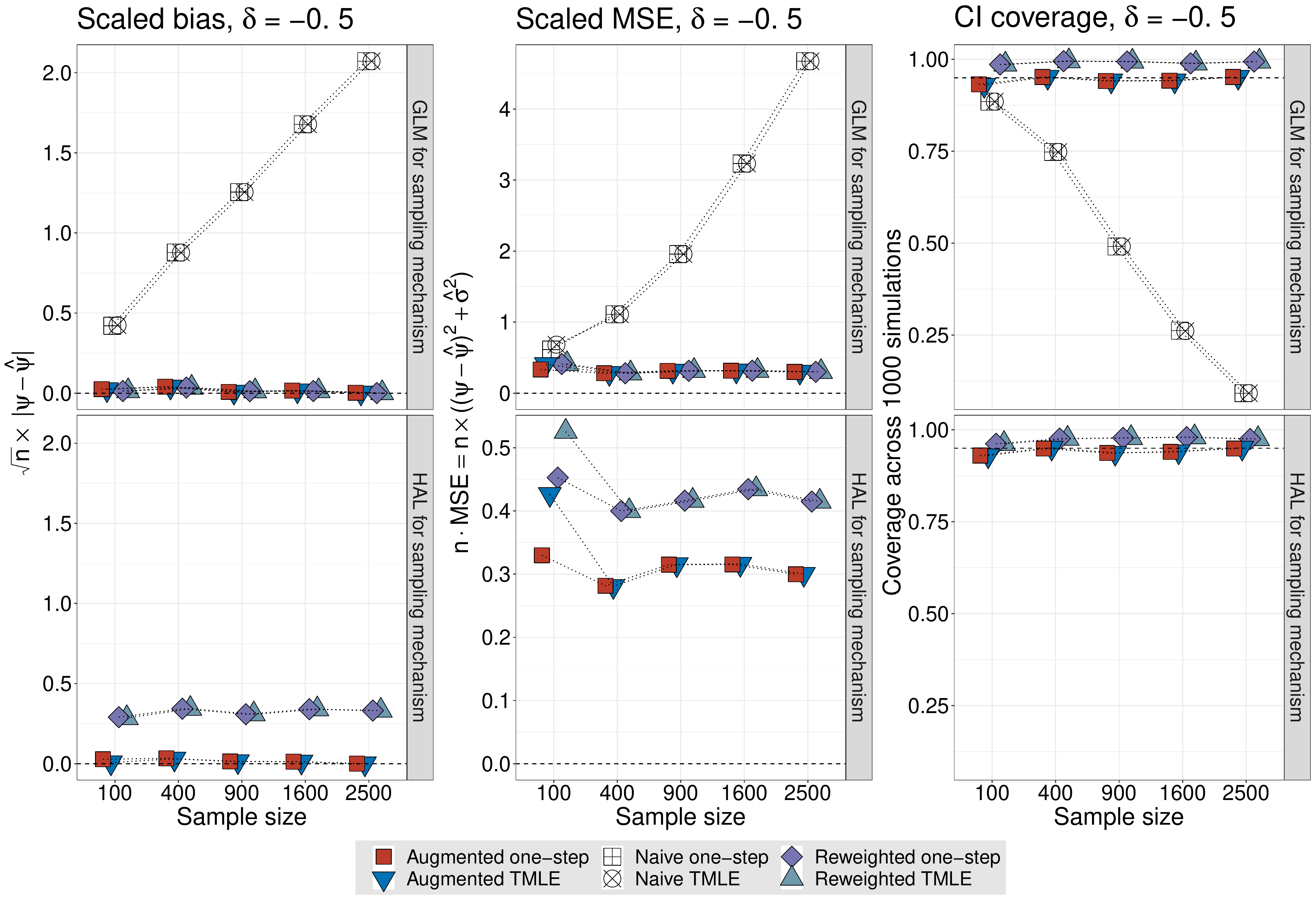}
  \caption{Results of numerical simulations comparing six estimation strategies
  for $\psi_{0,\delta}$ for $\delta = -0.5$.}
  \label{fig:simple_sim_delta_downshift}
\end{figure}

In the results reported in section~\ref{main:simple_sim}, for our first
simulation study under the shift $\delta = 0.5$, we noted excellent performance
of our proposed estimator variants under several standard metrics, including
$\sqrt{n}$-bias, $n$-MSE, and the coverage of confidence intervals. To further
assess the quality of performance of our augmented estimators, we examine the
same six estimator variants under the shifts $\delta \in \{-0.5, 0\}$. Details
of the simulation study have been previously described in
section~\ref{main:simple_sim}. As before, results are reported based on
aggregation across $1000$ repetitions. The results of these numerical
investigations are reported in Figures~\ref{fig:simple_sim_delta_noshift}
and~\ref{fig:simple_sim_delta_downshift}.

Both the reweighted estimators of~\citet{rose2011targeted2sd} and our augmented
estimators display the expected level of performance when the sampling mechanism
is estimated via a correctly specified parametric model, standing out against
the performance of their unadjusted counterparts. As in the case of $\delta
= 0.5$, the reweighted estimators display coverage exceeding the desired 95\%
level, while their augmented analogs cover at exactly the nominal level. This
suggests again that the reweighted estimators exhibit an inflated variance
relative to that of our augmented estimators. The lower panel of each figure
visualizes differences in the performance of the estimators when the sampling
mechanism is estimated flexibly via HAL. These results reveal a significant
discrepancy in the performance of the reweighted and augmented estimators, with
our augmented one-step and TML estimators outperforming their reweighted
counterparts in terms of $\sqrt{n}$-bias, $n$-MSE, and coverage. In this second
case, we note that the TML estimators appear to exhibit a small degree of
estimation instability at $n = 100$, displaying larger $n$-MSE in both the
reweighted and augmented variants. We conjecture that this performance is due to
an instability induced by the targeting procedure --- in practice, this could be
ameliorated by alterations to the convergence criterion. Overall, the results of
these numerical investigations do not differ substantially from those presented
in section~\ref{main:simple_sim}, establishing that our augmented estimators
improve upon alternative estimation strategies when nonparametric estimation of
$g_{0,C}$ is performed.

\subsection{Simulation \#2: Comparing estimators in a scenario inspired by HVTN
505, without effect of treatment}\label{sim2_supp}

\begin{figure}[H]
  \centering
  \includegraphics[scale=0.33]{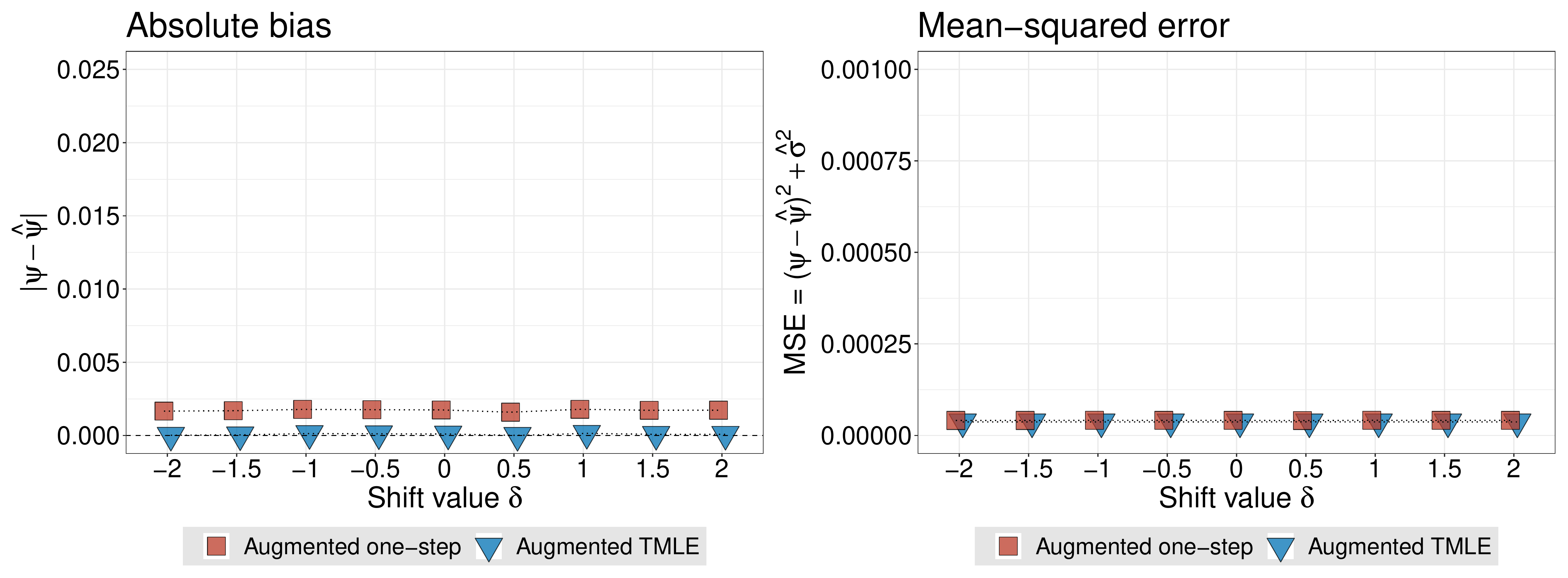}
  \caption{Results of numerical simulations comparing our two proposed
     estimators of $\psi_{0,\delta}$ for a grid of $\delta$, across 1000 Monte
     Carlo simulations at $n = 1400$.}
 \label{fig:hvtn_sim_null}
\end{figure}

To further assess the performance of our augmented estimators in a scenario like
the HVTN 505 trial, we replace the structural equation for the outcome $Y$ with
a draw from the Bernoulli distribution parameterized as $\bern (p = \expit(-2.8
- 0.0013 W_1 - 0.0016 W_2 + 0.0678 W_3 + 0.039 W_4))$ to achieve a setting in
which there is no effect of the treatment on the outcome. In this case, the
outcome process is relatively rare, with $\prob(Y = 1 \mid A, W) \approx
0.053$. The structural equations for other components of the observed data $O$
are unaltered. In this setting, the true parameter value was approximated to
be $\psi_{0,\delta} = 0.0526$ regardless of the value of $\delta$. We present
the results of assessing our augmented estimators in
Figure~\ref{fig:hvtn_sim_null}.

As the ground truth of the effect of shifting the post-vaccination activity of
the CD4+ and CD8+ immunogenic markers on the risk of HIV-1 infection is unknown
in the HVTN 505 trial, we assess our estimators in this scenario so as to be
sure of there performance when no effect of treatment is present. Estimation of
all nuisance parameters is performed using exactly the same techniques outlined
previously in section~\ref{main:hvtn_sim}.

In terms of bias and MSE, our proposed estimators display adequate performance:
the augmented one-step and TML estimators are effectively indistinguishable in
terms of MSE, while, in terms of bias, the augmented TML estimator appears to
outperform the one-step estimator uniformly across all values of $\delta$,
likely on account of the enhanced finite-sample performance of TML
estimators~\citep{vdl2011targeted}. As with the numerical investigations
presented in section~\ref{main:hvtn_sim}, these results suggest that our
augmented estimators will perform well enough to allow accurate estimation of
$\psi_{0,\delta}$ when applied to data from the HVTN 505 trial.

\section{Super Learner Ensemble Models for $\overline{Q}_{n,Y}$ in the HVTN 505
Data Analysis}\label{hvtn_sl_details}

As noted in section~\ref{main:application}, for both the CD4+ and CD8+ analyses,
the estimated outcome mechanism $\overline{Q}_{n,Y}$ was constructed from an
ensemble model based on the super learner algorithm~\citep{vdl2007super}. The
cross-validation selector that forms the basis of the super learner has been
shown to exhibit unique theoretical guarantees, including asymptotic equivalence
to the oracle selector~\citep{vdl2004asymptotic, vdl2006cross, vdl2006oracle},
that make its use preferable over other ensemble learning approaches.

A rich library of candidate classification algorithms was considered in the
super learner ensemble model for $\overline{Q}_{n,Y}$. These included
$\ell_1$-penalized lasso regression~\citep{tibshirani1996regression,
friedman2009glmnet}; $\ell_2$-penalized ridge regression
\citep{tikhonov1977solutions,hoerl1970ridge,friedman2009glmnet}; three elastic
net regressions weighting the $\ell_1$ penalty at $\alpha \in \{0.25, 0.50,
0.75\}$ and the $\ell_2$ penaly at
$(1-\alpha)$~\citep{zou2003regression,friedman2009glmnet}; random
forests~\citep{breiman2001random} with 50, 100, and 500 trees based on its
implementation in the \texttt{ranger} \texttt{R}
package~\citep{wright2017ranger}; extreme gradient boosted trees with 20, 50,
100, and 300 fitting iterations~\citep{chen2016xgboost}; multivariate adaptive
polynomial spline regression~\citep{kooperberg1997polychotomous,stone1994use};
multivariate adaptive regression splines~\citep{friedman1991multivariate};
generalized linear models with Bayesian priors on parameters; a multilayer
perceptron~\citep{rosenblatt1961principles}; and the highly adaptive
lasso~\citep{vdl2017generally,benkeser2016highly,coyle2019hal9001}. The
implementation of the super learner algorithm in the \texttt{sl3} \texttt{R}
package~\citep{coyle2020sl3} was used, and weights assigned to each learning
algorithm by the super learner are given in Tables~\ref{tab:cd4_sl_coefs}
and~\ref{tab:cd8_sl_coefs} for the CD4+ and CD8+ analyses, respectively.

\begin{table}[H]
\caption{\label{tab:cd4_sl_coefs} Weights and risk estimates assigned to each
  individual learning algorithm in the ensemble model for $\overline{Q}_{n,Y}$
  used in the reported re-analysis of the CD4+ immunogenic marker.}
\centering
\begin{tabular}[t]{l|r|r|r|r}
\hline
Learner & Weight & Min. Fold Risk & Mean CV-Risk & Max. Fold Risk\\
\hline
Ridge ($\ell_2$ penalized) & 0.147 & 0.015 & 0.036 & 0.052\\
\hline
Lasso ($\ell_1$ penalized) & 0.000 & 0.015 & 0.036 & 0.052\\
\hline
Elastic net ($\alpha = 0.25$) & 0.116 & 0.015 & 0.036 & 0.051\\
\hline
Elastic net ($\alpha = 0.50$) & 0.000 & 0.015 & 0.036 & 0.051\\
\hline
Elastic net ($\alpha = 0.75$) & 0.181 & 0.015 & 0.036 & 0.051\\
\hline
Random forest (50 trees) & 0.000 & 0.062 & 0.097 & 0.173\\
\hline
Random forest (100 trees) & 0.000 & 0.050 & 0.093 & 0.153\\
\hline
Random forest (500 trees) & 0.000 & 0.061 & 0.093 & 0.158\\
\hline
xgboost(20 iterations) & 0.000 & 0.012 & 0.072 & 0.165\\
\hline
xgboost(50 iterations) & 0.000 & 0.012 & 0.080 & 0.181\\
\hline
xgboost(100 iterations) & 0.010 & 0.013 & 0.088 & 0.192\\
\hline
xgboost(500 iterations) & 0.000 & 0.012 & 0.098 & 0.204\\
\hline
Highly adaptive lasso (default) & 0.014 & 0.064 & 0.074 & 0.084\\
\hline
Highly adaptive lasso (custom) & 0.000 & 0.066 & 0.075 & 0.084\\
\hline
Multivariate adaptive regression splines & 0.000 & 0.050 & 0.086 & 0.131\\
\hline
Polynomial spline regression & 0.208 & 0.015 & 0.036 & 0.053\\
\hline
Multilayer perceptron network & 0.000 & 0.015 & 0.037 & 0.055\\
\hline
GLM with Bayesian priors & 0.323 & 0.015 & 0.036 & 0.051\\
\hline
Super Learner & 1.000 & --- & --- & ---\\
\hline
\end{tabular}
\end{table}

In the super learner ensemble model for the outcome regression in the CD4+
analysis, the three best learning algorithms were a GLM with Bayesian priors on
parameter estimates, a polynomial spline regression model, and an elastic net
regression model that favored the $\ell_1$ (lasso) penalty over the $\ell_2$
(ridge) penalty. Another variant of elastic net regression, which favored the
$\ell_2$ penalty over the $\ell_1$ penalty, and ridge regression were also given
nontrivial weights in the ensemble model. A variant of extreme gradient boosting
and the highly adaptive lasso received low weights, while all other candidate
algorithms in the library were assigned weights of zero.

\begin{table}[H]
\caption{\label{tab:cd8_sl_coefs} Weights and risk estimates assigned to each
  individual learning algorithm in the ensemble model for $\overline{Q}_{n,Y}$
  used in the reported re-analysis of the CD8+ immunogenic marker.}
\centering
\begin{tabular}[t]{l|r|r|r|r}
\hline
Learner & Weight & Min. Fold Risk & Mean CV-Risk & Max. Fold Risk\\
\hline
Ridge ($\ell_2$ penalized) & 0.161 & 0.013 & 0.037 & 0.075\\
\hline
Lasso ($\ell_1$ penalized) & 0.161 & 0.011 & 0.037 & 0.075\\
\hline
Elastic net ($\alpha = 0.25$) & 0.003 & 0.012 & 0.037 & 0.075\\
\hline
Elastic net ($\alpha = 0.50$) & 0.000 & 0.014 & 0.037 & 0.076\\
\hline
Elastic net ($\alpha = 0.75$) & 0.131 & 0.014 & 0.037 & 0.074\\
\hline
Random forest (50 trees) & 0.090 & 0.053 & 0.088 & 0.129\\
\hline
Random forest (100 trees) & 0.055 & 0.048 & 0.089 & 0.136\\
\hline
Random forest (500 trees) & 0.119 & 0.049 & 0.088 & 0.127\\
\hline
xgboost(20 iterations) & 0.000 & 0.043 & 0.064 & 0.112\\
\hline
xgboost(50 iterations) & 0.000 & 0.036 & 0.074 & 0.128\\
\hline
xgboost (100 iterations & 0.000 & 0.031 & 0.076 & 0.134\\
\hline
xgboost (300 iterations) & 0.000 & 0.029 & 0.086 & 0.146\\
\hline
Highly adaptive lasso (default) & 0.000 & 0.046 & 0.078 & 0.115\\
\hline
Highly adaptive lasso (custom) & 0.000 & 0.044 & 0.078 & 0.132\\
\hline
Multivariate adaptive regression splines & 0.000 & 0.029 & 0.100 & 0.159\\
\hline
Polynomial spline regression & 0.000 & 0.015 & 0.040 & 0.080\\
\hline
Multilayer perceptron network & 0.067 & 0.007 & 0.040 & 0.085\\
\hline
GLM with Bayesian priors & 0.214 & 0.016 & 0.037 & 0.073\\
\hline
Super Learner & 1.000 & --- & --- & ---\\
\hline
\end{tabular}
\end{table}

In the CD8+ analysis, the three best learning algorithms, chosen by the super
learner ensemble model for the outcome regression, were a GLM with Bayesian
priors on parameter estimates, ridge regression, and lasso regression. Other
algorithms that were assigned nontrivial weights included an elastic net
regression model that favored the $\ell_1$ (lasso) penalty over the $\ell_2$
(ridge) penalty and a random forest with 500 trees. A variant of elastic net
regression, random forests with 50 and 100 trees, and a multilayer perceptron
all received low weights, while all other candidate algorithms in the library
were assigned weights of zero.

\bibliographystyle{biom}
\bibliography{refs}